\documentclass[useAMS,usenatbib]{mn2e}

\usepackage{aas_macros}

\usepackage[multiple]{footmisc}
\usepackage{rotating}
\usepackage{fixltx2e}   

\title[N-body Simulations of the Magellanic Stream]{
  N-body Simulations of the Magellanic Stream
}

\author[Connors, Kawata \& Gibson]
{
  Tim W. Connors$^{1}$\thanks{E-mail: tconnors@astro.swin.edu.au},
  Daisuke Kawata$^{1,2}$,
  Brad K. Gibson$^{3}$\\
  $^1$Centre for Astrophysics \& Supercomputing, 
  Swinburne University, Hawthorn, VIC 3122, Australia\\
  $^2$The Observatories of the Carnegie Institution of Washington,
  813 Santa Barbara Street, Pasadena, CA 91101, USA\\
  $^3$Centre for Astrophysics, University of Central
  Lancashire, Preston, PR1 2HE, United Kingdom
}
\begin{document}

\newcommand{\HI}{\ifmmode{\rm HI}\else{\mbox{H\,{\sc{i}}}}\fi}
\newcommand{\e}[1]{\ensuremath{\times 10^{#1}}}
\newcommand{\msun}{\ensuremath{{\rm \,M}_{\sun}}}
\newcommand{\kms}{\ensuremath{{\rm \,km\,s}^{-1}}}
\newcommand{\Gyr}{\ensuremath{{\rm \,Gyr}}}
\newcommand{\Gyrs}{\ensuremath{{\rm \,Gyrs}}}
\newcommand{\kpc}{\ensuremath{{\rm \,kpc}}}
\newcommand{\pc}{\ensuremath{{\rm \,pc}}}
\newcommand{\kmsns}{\ensuremath{{\rm km\,s}^{-1}}}
\newcommand{\msunns}{\ensuremath{{\rm M}_{\sun}}}
\newcommand{\kpcns}{\ensuremath{{\rm kpc}}}
\newcommand{\Jykms}{\ensuremath{{\rm \,Jy\,\kms{}}}}
\newcommand{\vlsr}{\ensuremath{{v}_{\rm LSR}}}
\newcommand{\vgsr}{\ensuremath{{v}_{\rm GSR}}}
\newcommand{\vms}{\ensuremath{{v}_{\rm MS}}}
\newcommand{\lMC}{\ensuremath{{\theta}_{\rm MC}}}
\newcommand{\simgt}{\ensuremath{\ga}}
\newcommand{\simlt}{\ensuremath{\la}}

\newcommand{\Fig}[1]{Fig.~\ref{fig:#1}}
\newcommand{\Figs}[1]{Figs.~\ref{fig:#1}}
\newcommand{\Tab}[1]{Table~\ref{tab:#1}}
\newcommand{\Sec}[1]{Section~\ref{sec:#1}}
\newcommand{\eqn}[1]{equation~(\ref{eqn:#1})}
\newcommand{\Eqn}[1]{Equation~(\ref{eqn:#1})}  
\newcommand{\eqnsII}[2]{equations~(\ref{eqn:#1}) and (\ref{eqn:#2})}
\newcommand{\EqnsII}[2]{Equations~(\ref{eqn:#1}) and (\ref{eqn:#2})}  

\newcommand{\code}[1]{{\mbox{\sc#1}}}
\newcommand{\GCD}{{\code{GCD+}}}
\newcommand{\galactICs}{\code{GalactICs}}
\newcommand{\pview}{{\code{PView}}}

\newcommand{\fixme}[1]{FIXME: {\it #1}}

\defcitealias{gn96}{GN96}
\defcitealias{yn03}{YN03}
\defcitealias{g99}{G99}
\defcitealias{gsf94}{GSF94}

\date{\today}

\pagerange{\pageref{firstpage}--\pageref{lastpage}}

\pubyear{2005}

\maketitle

\label{firstpage}

\begin{abstract}
  A suite of high-resolution N-body simulations of the Magellanic
  Clouds -- Milky Way system are presented and compared directly with
  newly available data from the \HI{} Parkes All-Sky Survey (HIPASS).
  We show that the interaction between Small and Large Magellanic
  Clouds results in both a spatial and kinematical bifurcation of both
  the Stream and the Leading Arm.  The spatial bifurcation of the
  Stream is readily apparent in the HIPASS data, and the kinematical
  bifurcation is also tentatively identified.  This bifurcation
  provides strong support for the tidal disruption origin for the
  Magellanic Stream.  A fiducial model for the Magellanic Clouds is
  presented upon completion of an extensive parameter survey of the
  potential orbital configurations of the Magellanic Clouds and the
  viable initial boundary conditions for the disc of the Small
  Magellanic Cloud.  The impact of the choice of these critical
  parameters upon the final configurations of the Stream and Leading
  Arm is detailed.
\end{abstract}


\section{Introduction}
\label{sec:intro}

The progressive collapse and merging associated with hierarchical
clustering within a cold dark matter cosmology, while dominated by
activity at early epochs (redshifts $z \sim 2$ -- 5; e.g.\
\citealp{mkh+02}), continues to the present-day and is readily
observable even in the local Universe.  In our own Milky Way (MW),
ongoing satellite disruption and accretion events include the
Sagittarius dwarf \citep{igi94}, the Canis Major dwarf
(\citealp{mic+05}, and references therein), and perhaps the most
spectacular of all, the Large (LMC) and Small (SMC) Magellanic Clouds
(MCs).  The disruption and accretion of the MCs is perhaps best
appreciated through the nearly circum-Galactic polar ring of gas --
the Magellanic Stream -- emanating from the Clouds
\citep{mcm74,pgs+98}.

Two primary, competing, scenarios have been postulated to explain the
origin of the Magellanic Stream (MS): (i) ram pressure stripping of
LMC and SMC gas due to the motion of the Clouds through the tenuous
coronal gas in the Galactic halo \citep{md94,mmm+05}.  This ``drag''
scenario faces difficulty in explaining the Leading Arm Feature (LAF)
observed by \citet{pgs+98}.  (ii) tidal disruption of the SMC
\citep{mf80}.  Because of its ability to simultaneously produce both
trailing and leading streams of gas, the ``tidal'' model has gathered
considerable support in the literature (\citealp{gn96}; \citealp{g99};
\citealp{yn03}, hereafter, \citetalias{gn96}, \citetalias{g99} and
\citetalias{yn03} respectively).

Constraints upon theoretical models of the Magellanic Stream have
improved dramatically with the release of the \HI{} Parkes All-Sky
Survey (HIPASS; \citealp{bsd+01}) dataset.  The extent and fine-scale
structure of the MS can now be appreciated to a level not previously
possible, in particular the unequivocal evidence for the existence of
the Leading Arm Feature \citep{pgs+98}.  Motivated by HIPASS, we have
initiated a program of high-resolution N-body modelling of the
Magellanic Stream aimed at de-constructing the temporal evolution of
the Magellanic Clouds -- Milky Way interaction.  Here, we use $\sim
30$ times higher resolution than previous models such as
\citetalias{gn96}, \citetalias{g99} and
\citetalias{yn03}\footnote{\citet{bc05} have presented high-resolution
  simulations of the Magellanic System, focusing exclusively on the
  internal dynamics and star formation history of the LMC by fixing
  the potentials of both the Milky Way and SMC - i.e.  predictions
  concerning the formation and evolution of the Magellanic Stream and
  Leading Arm were not features of their work.  The mass resolution
  (softening length) employed in our work is approximately a factor of
  ten (two) greater than that of \citet{bc05}.}, giving us a \HI{}
mass resolution of $\sim 5600 \msun$ and a \HI{} flux resolution of $7
\Jykms$.  In addition, we construct a detailed \HI{} map, to compare
with the HIPASS data directly and quantitatively.  The combination of
a higher resolution simulation and the simulated \HI{} map enables us
to argue about more detailed features in the MS and the LAF.  As a
result, this paper shows more convincingly that the observed LAF and
MS can be produced by the leading and trailing streams of the SMC,
induced by tidal interactions with the MW and LMC.  Our fiducial model
is found after an extensive parameter survey of the potential orbital
configurations of the MCs and the different initial condition of the
SMC.  Based on the parameter survey, we also demonstrate how the final
features of the MS and the LAF are sensitive to the initial
configuration of the SMC.  Our new high-resolution simulations also
reveal that the tidal interactions create spatial and kinematical
bifurcation in the MS and LAF.  We present this prediction based on
our simulations, however we also demonstrate that the existing HIPASS
data show such bifurcations.

The work described here (Paper~I) is the first in a series of papers
aimed at providing a definitive model for the MS and the LAF.  Here,
we concentrate solely upon the effects of gravity, describing the SMC
disc with collisionless particles, and ignore the baryon physics
(including hydrodynamics, star formation, energy feedback, and
chemical enrichment).  In Paper~II, the effects of baryonic physics
will be detailed.

The layout of this paper is as follows: in \Sec{simulation} we provide
a description of the suite of simulations generated to date.  In
\Sec{fiducial}, we show the results from our best model, and compare
them with the observational data.  In \Sec{paramdep}, we demonstrate
how the final features of the MS and LAF depend upon the orbits of the
MCs and the initial properties of the SMC disc.  Finally, in
\Sec{summary}, the discussion and future directions for our work are
presented.

\section{Numerical Simulation}
\label{sec:simulation}

The framework upon which our simulations are based parallels that
described by \citetalias{gn96} and \citetalias{yn03}.  The MW and LMC
are taken to be fixed potentials, while the SMC is treated as an
ensemble of self-gravitating particles, in recognition of the fact
that the MS is thought to originate from the tidal disruption of the
SMC disc (e.g.\ \citealp{gsf94}, hereafter \citetalias{gsf94};
\citetalias{gn96}; \citealp{mkg02}; \citetalias{yn03}; but see also
\citealp{mmm+05}).  The orbits of the MW and LMC with respect to the
SMC are pre-calculated, the procedure for which is outlined in
\Sec{simulation/orb} and \citetalias{gsf94}.  The initial boundary
conditions for the ``live'' SMC model is described in
\Sec{simulation/smc}, and its evolution explored in
\Sec{simulation/evolve}.

\subsection{The orbits of the LMC and SMC}
\label{sec:simulation/orb}

We assume a spherically symmetric potential for the MW halo.  The MCs
sample a volume of the halo where the gravitational potential is
insensitive to both the expected central cusp and the outer regions
where the density profile may be steeper than an isothermal profile.
We hence assume a constant rotational velocity of \mbox{$V_{\rm
    c}=220\kms$} within the MW halo.  Thus, the potential is described
by
\begin{equation}
  \phi_{\rm G}(r) =
  -V_{\rm c}^2\ln r
  \label{eqn:potMW}
\end{equation}
with a mass enclosed within $r \kpc$ of
\begin{equation}
  M_{\rm G}(<r) = 5.6\e{11}\left(\frac{V_{\rm c}}{220
      \kms}\right)^2\left(\frac{r}{50\kpc}\right)\msun .
  \label{eqn:massfrompot}
\end{equation}
We assume that there is little disc contribution to the potential at
the typical Galactocentric distances of the LMC and SMC ($\sim 50
\kpc$).  The halos of both the MW and LMC are assumed to be invariant
for the duration of the simulation ($2.5 \Gyr$ -- i.e., a lookback
time corresponding to redshift $z \sim 0.2$).  Plummer potentials are
adopted for both the LMC and SMC -- i.e.,
\begin{equation}
  \phi_{\rm L,S}({\bf r}) = G M_{\rm L,S}/\left[\left({\bf r}
      - {\bf r}_{\rm L,S}\right)^2 + K_{\rm L,S}^2\right]^{1/2} ,
  \label{eqn:potL,SMC}
\end{equation}
where ${\bf r}_{\rm L,S}$ are the positions of the Clouds relative to
the MW centre, and $K_{\rm L,S}$ are the core radii, set to $3$ and
$2\kpc$ for the LMC and SMC, respectively.  In the fiducial model, we
assume a constant mass of $M_{\rm L}=2\e{10}\msun$ for the LMC, and
$M_{\rm S}=3\e9\msun$ for the SMC.  

Adopting these parameters, we next backwards integrate the orbits of
the SMC and LMC, using as boundary conditions the current tangential
and radial velocities and positions of the Clouds listed in
\Tab{curvels} (see also \citealp{mf80}, \citetalias{gsf94},
\citetalias{gn96}, \citealp{ljk95} and \citealp{bcb+04}).  The values
we adopted for our work were chosen to match those of
\citetalias{gn96}, and are consistent with the extant literature.

\begin{table}
  \caption[Orbital parameters for the SMC and LMC]{Present-day orbital
    parameters for the SMC and LMC from the literature and values we
    have adopted in this work.
  }
  \begin{minipage}[P]{7.5cm}
    \renewcommand{\thempfootnote}{\arabic{mpfootnote}}%
    \begin{tabular}{lllll}
      Parameter & \multicolumn{2}{c}{LMC} & \multicolumn{2}{c}{SMC} \\
      & Literature & This & Literature & This \\
      &            & work &            & work \\
      \hline
      $v_{\rm r,GSR}$\footnote{Radial velocity
        in the Galactic Standard of Rest frame ($\kmsns$)} & 
      $84 \pm 7$ \footnote{\citet{vah+02}\label{ftn:vah+02}} & 
      $80.1$ & $\sim 7$ \footnote{\citet{hsa89}}\footnote{\citetalias{gsf94}} &
      $7.1$ 
      \\
      $v_{\rm t,GSR}$\footnote{Tangential velocity
        in the Galactic Standard of Rest frame ($\kmsns$)} & 
      $281 \pm 41$ \footref{ftn:vah+02} &
      $287$ & $200\pm100$ \footnote{\cite{ljk95}} & $255$
      \\
      $l$\footnote{Galactic longitude} &
      $280.46\degr$ \footnote{\citet{t88}\label{ftn:t88}} & $280.46\degr$ &
      $302.79\degr$ \footref{ftn:t88} & $302.79\degr$ 
      \\
      $b$\footnote{Galactic latitude} & 
      $-32.89\degr$ \footref{ftn:t88} & $-32.89\degr$ & 
      $-44.30\degr$ \footref{ftn:t88} & $-44.30\degr$ 
      \\
      $d$\footnote{Distance ($\kpcns$)} & $49.43$ \footnote{\citet{fw87}\label{ftn:fw87}} 
      & $49.43$ & $57.02$\footref{ftn:fw87} & $57.02$ \\
    \end{tabular}
  \end{minipage}
  \label{tab:curvels}
\end{table}

The equations of motion of the Clouds about the stationary MW are
\begin{equation}
  \ddot{\bf r}_{\rm L} =
  \frac{\partial}{\partial {\bf r}_{\rm L}} \left[\phi_{\rm
      S}\left(\left|{\bf r}_{\rm L} - {\bf
    r}_{\rm S}\right|\right)+\phi_{\rm G}\left(\left|{\bf r}_{\rm
    L}\right|\right)\right] + {\bf F}_{\rm L}/M_{\rm L}
   \label{eqn:orbaccl}
\end{equation}
and
\begin{equation}
  \ddot{\bf r}_{\rm S} =
  \frac{\partial}{\partial {\bf r}_{\rm S}} \left[\phi_{\rm
      L}\left(\left|{\bf r}_{\rm S} - {\bf
    r}_{\rm L}\right|\right)+\phi_{\rm G}\left(\left|{\bf r}_{\rm
    S}\right|\right)\right] + {\bf F}_{\rm S}/M_{\rm S} ,
   \label{eqn:orbaccs}
\end{equation}
where the potentials $\phi_{\rm L}$, $\phi_{\rm S}$ and $\phi_{\rm G}$
refer to the Large and Small Magellanic Clouds, and the Galaxy,
respectively \citep{mf80}.  Since we do not model the MW as live
particles, dynamical friction is modelled as per \citetalias{gsf94}.
${\bf F}_{\rm L}$ and ${\bf F}_{\rm S}$ are the dampening force
between the Galaxy and each of the Clouds:
\begin{eqnarray}
  \lefteqn{{\bf F}_{\rm L,S} =  \frac{-G M_{\rm L,S}}{\left|{\bf
          r}_{\rm L,S}\right|^2} 
    \ln\left(\Lambda\right)
    \frac{\dot{\bf r}_{\rm L,S}}{\left|\dot{\bf r}_{\rm L,S}\right|}} & \nonumber \\
  & 
  \mbox{} \times
  \left\{
    \frac{2 \left[
        \int_{0}^{x}{
          \exp(-y^2) dy - \exp(-x^2)x
        }
      \right]}
    {\pi^{1/2}x^2}
  \right\}_{x=
    \left[\left|\dot{\bf r}_{\rm L,S}\right|^2/V_{\rm c}\right]^{1/2}} ,
  \label{eqn:dynfric}
\end{eqnarray}
where $\ln (\Lambda)$ is the Coulomb logarithm.  

Although \eqn{dynfric} is a simple analytical formulation for
dynamical friction that fails to model accurately the orbit of the
Clouds for more than $5 \Gyr$, 
according to recent studies \citep{hfm03,jp05}, it gives a good
approximation to the predictions of full N-body simulations over the
$\sim 2.5 \Gyr$ period we focus on in this paper.  We note also, the
strong effect the Coulomb logarithm has on the orbits.  \citet{hfm03}
suggest that the $\ln (\Lambda)$ value of 3 advocated by \citet{bt87}
for the LMC is too large, causing the system to evolve too fast.
Thus, for this paper, we set $\ln (\Lambda) = 1$ to enable direct
comparison with the previous works.

\begin{figure*}
  \begin{center}
    \includegraphics[width=8cm]{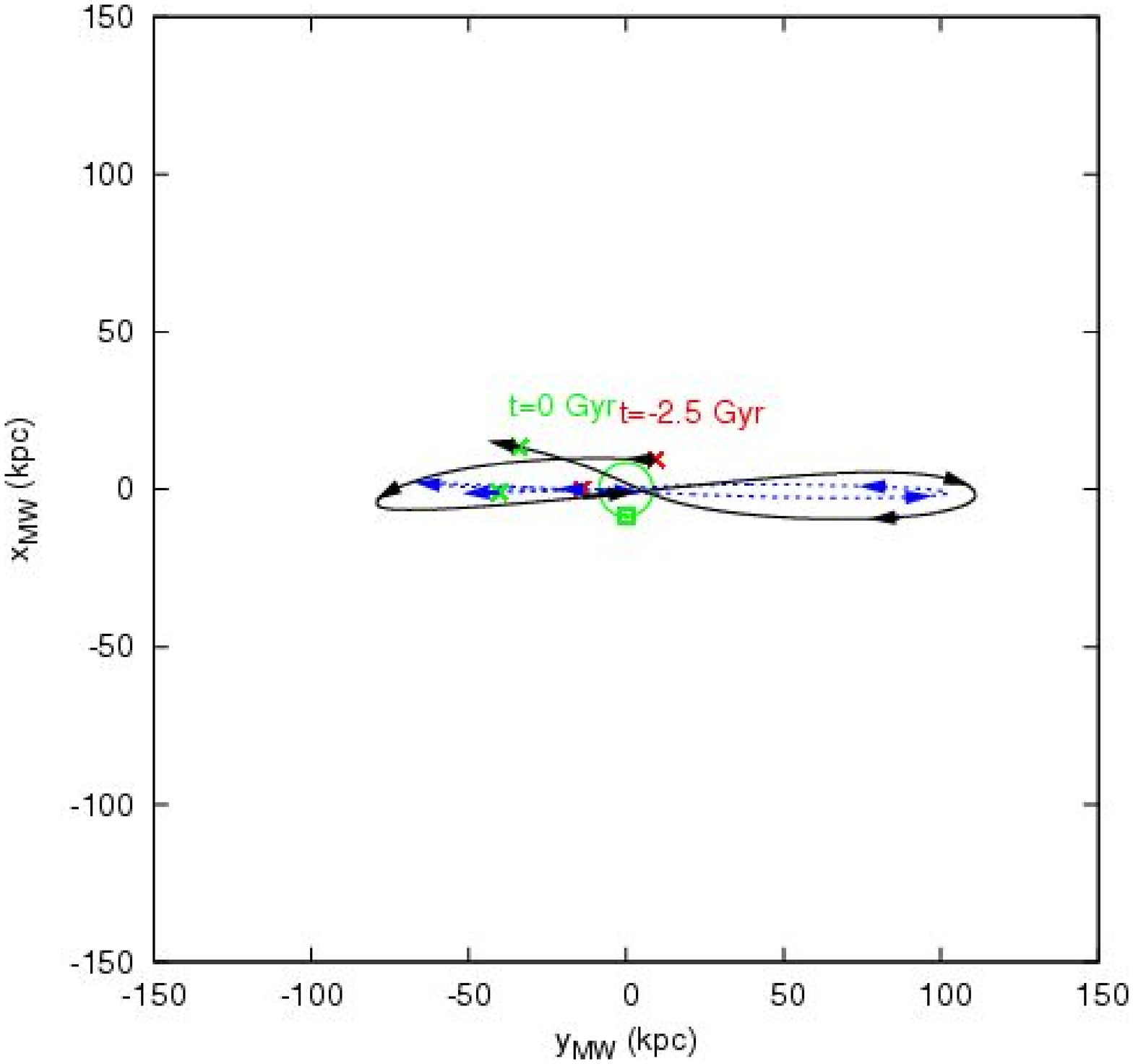}
    \includegraphics[width=8cm]{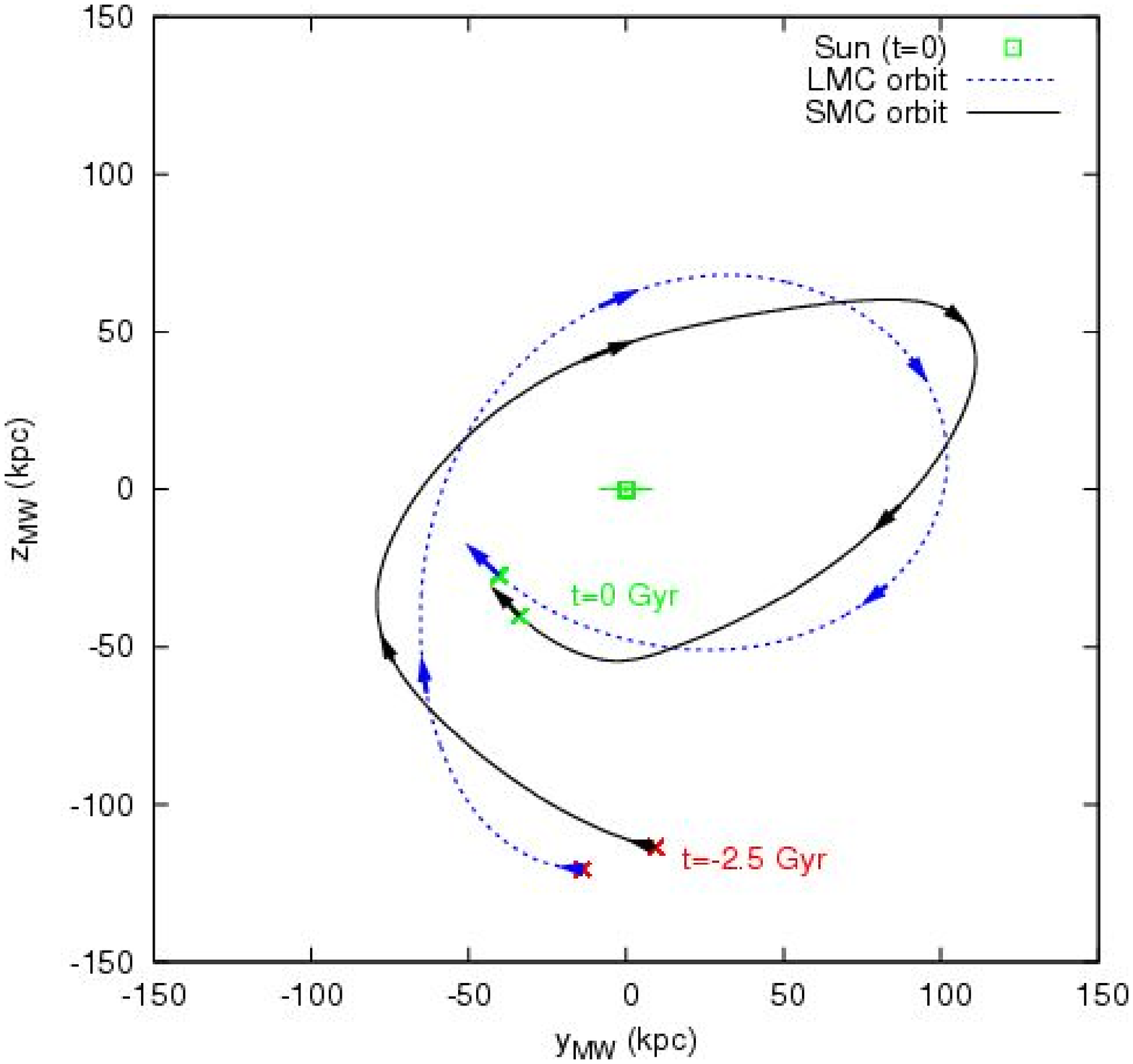}
  \end{center}
  \caption[Three dimensional orbit of the LMC and SMC around the
  Galaxy]{The three dimensional orbit of the SMC (black line) and the
    LMC (blue dashed line) around the Galaxy.  The rotation axis of
    the MW is assumed to be the $z$-axis, and the disc plane is
    centred on $z=0$.  The orbits, which are derived by backwards
    integrating to $T=-2.5 \Gyr$, are plotted with arrows indicating
    each $0.5 \Gyr$ between $T=0$ and $-2.5 \Gyr$.  The green crosses
    denote the current position of the SMC and LMC.  The Solar radius
    at $8.5\kpc$ is also shown, and the current position of the Sun at
    $T=0$ is marked by the open square.}
  \label{fig:xyzorbit} 
\end{figure*}

\begin{figure}
  \begin{center}
    \includegraphics[width=8cm]{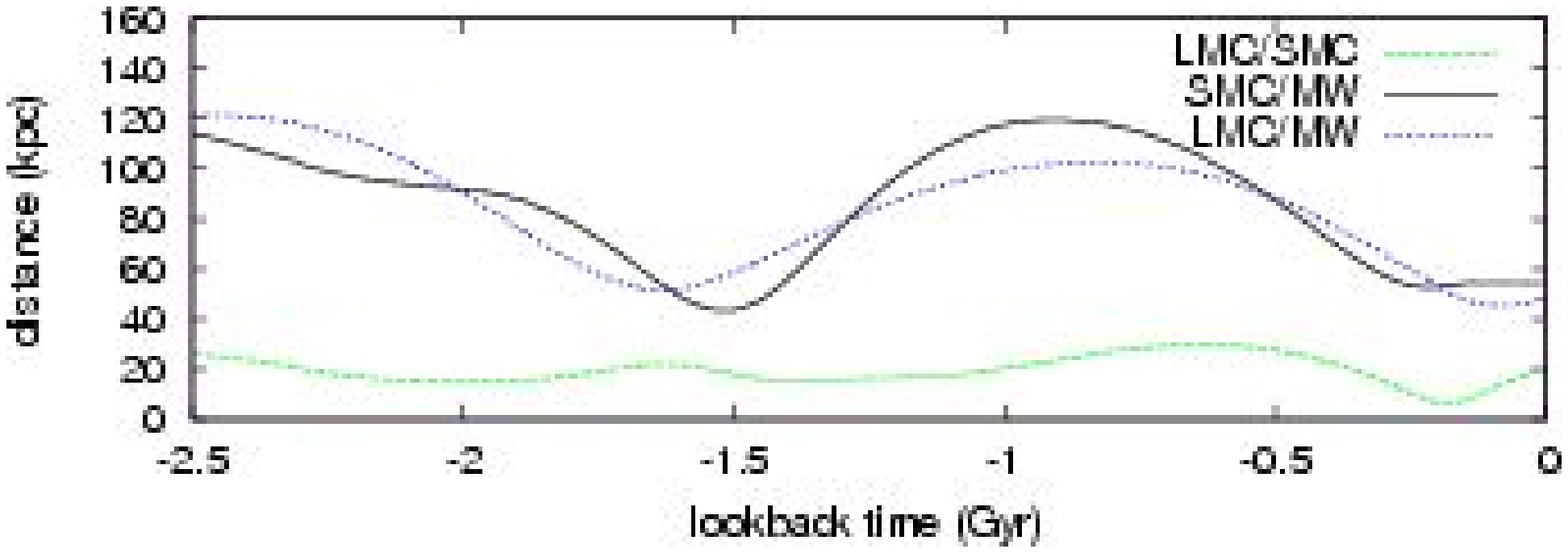}
    \includegraphics[width=8cm]{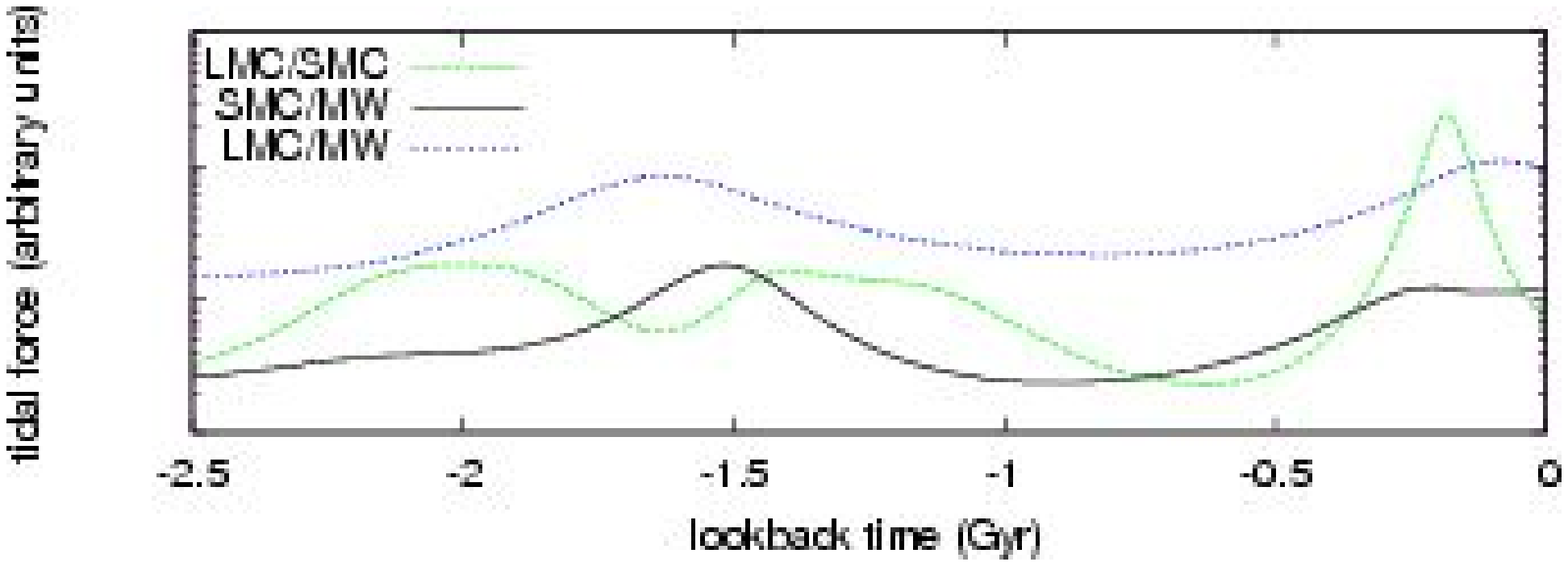}
  \end{center}
  \caption[Distance and tidal forces between the three bodies over
  time]{Distance (upper) and approximate tidal forces (lower; on a
    logarithmic scale in arbitrary units) between the three
    interacting bodies in our fiducial model.}
  \label{fig:disttidalorbit} 
\end{figure}

Our fiducial model is initiated at a lookback time of $2.5 \Gyrs$ when
the Clouds are at apo-Galacticon (with the tidal forces between all
three bodies being minimised).  The subsequent orbit appears in
\Fig{xyzorbit}, and the distance between the Clouds and Galaxy is
shown in \Fig{disttidalorbit}.

\begin{table*}
  \caption[Fiducial SMC initial model parameters]{Fiducial SMC model
    parameters (see \citealp{kd95} for details of parameters).}
  \begin{minipage}[P]{18cm}
    \renewcommand{\thempfootnote}{\arabic{mpfootnote}}%
    \begin{tabular}{lllllllllllll}
      & \multicolumn{6}{c}{Disc} & 
      & \multicolumn{5}{c}{Halo} \\
      & $M_{\rm d}$ 
      & $R_{\rm d}$ & $R_{\rm t}$
      & $z_{\rm d}$ & $\delta R_{\rm out}$
      & $\sigma_{\rm R,0}$ &
      & $\Psi_0$ & $\sigma_0$ & $q$ 
      & $C$ 
      & $R_{\rm a}$
      \\
      & ($\msunns$) & ($\kpcns$) & ($\kpcns$) & ($\kpcns$) & ($\kpcns$) & ($\kmsns$) &
      & (${\rm km}^2 {\rm s}^{-2}$)  & &  &  & ($\kpcns$) \\
      \hline
      & $2.47\e9$ & $3.5$ & $7.0$ & $0.35$ & $1.1$ & $35$ &
      & $-5.72\e3$ & $29$ & $1.0$ & $0.1$ & $0.96$
   \end{tabular}
  \end{minipage}
  \label{tab:galactics-in}
\end{table*}

\subsection{SMC model}
\label{sec:simulation/smc}

The initial dynamical configuration for the SMC was constructed using
\galactICs{} \citep{kd95}.  We employed a bulgeless equilibrium model
with a truncated exponential disc, and a dark matter King-profile
halo.  The relevant \galactICs{} parameters for the fiducial SMC model
appear in \Tab{galactics-in} (a detailed explanation of the individual
parameters is provided by \citealp{kd95}).  The disc possesses an
exponential profile with scale radius $R_{\rm d}=3.5 \kpc$, smoothly
truncated beyond $R_{\rm t}=7 \kpc$ (with 95 per~cent of both the disc
and halo masses being within $r_{\rm d,95} \sim r_{\rm h,95} \sim
7\kpc$), to give a disc with total radial extent (where the face-on
surface density of the disc reaches $1 \msun \pc^{-2}$) of $7.5\kpc$
(compared to the halo with radial extent $\sim 14 \kpc$).  The
rotation curve peaks at $\sim 2 \kpc$ with a velocity of $\sim 45
\kms$, and turns over to become approximately constant, giving a total
SMC mass of $3\e9\msun$, with the disc mass being $1.5\e9\msun$.  The
central velocity dispersion of the disc was chosen to be near the
current \HI{} velocity dispersion of the observed SMC
(\Fig{skymom12}), and the dark matter halo velocity dispersion is
similar to the value of $\sim 25 \kms$ obtained from observations of
the stellar halo carbon stars and planetary nebulae
\citep{dlf+85,hsa89,hcm+97}.  The Toomre $Q$-parameter at the disc
half-mass radius is $Q=1.4$.

Our fiducial SMC model assumes a somewhat different scale length and
total extent for the SMC disc when compared with earlier studies
(\citetalias{gsf94}; \citetalias{gn96}; \citetalias{yn03}).  We should
stress that one cannot simply take as initial conditions, observed
parameters of the SMC at the {\it present day}, since the SMC has
(obviously) been significantly disturbed by interactions with the MW
and LMC; we instead use initial parameters consistent with other
relatively isolated dwarf disc galaxies (and integrate forward to
ensure the final characteristics is consistent with that observed
today).  Dwarfs with \HI{} mass $1.3$ -- $2.0\e9 \msun$ (comparable to
the SMC) have \HI{} disc scale lengths ranging from $1.6$ to
$4.4\kpc$, and smaller optical (i.e.\ stellar) scale lengths of $0.9$
to $2.7\kpc$ \citep{svv+02}.  These same galaxies have \HI{} discs
whose face-on corrected \HI{} density drops to $1 \msun \pc^{-2}$ at a
radius between $10$ to $12 \kpc$.

The reasons for the larger truncation radius compared to previous work
are twofold.  First, we suggest the tidal radius of the SMC disc is
larger than the $5\kpc$ proposed by \citetalias{gn96}.  Several Gyrs
ago, the MW was likely to have been slightly less massive (secular
halo growth), the SMC slightly more massive (less tidal disruption),
and the perigalactic distance of the SMC was larger.  In the estimated
orbit of the SMC, at a lookback time of $2.5 \Gyr$
(\Fig{disttidalorbit}), the perigalactic distance $r_{\rm p}$ was
$60\kpc$, and the apogalactic distance $r_{\rm a}$ was $120\kpc$.
However, the MW mass enclosed within the larger SMC orbit has then
increased.  The tidal radius of the SMC is \citep{fl83}
\begin{equation}
  r_{\rm t} = r_{\rm p}\left[\frac{M_{\rm
        S}}{(3+e)M_{\rm G}(r_{\rm p})}\right]^{1/3},
  \label{eqn:tidalrad}
\end{equation}
where the eccentricity $e=(1-r_{\rm p}^2/r_{\rm a}^2)^{0.5}$, and thus
$r_{\rm t} = 6.3\kpc$ when $r_{\rm p}=60\kpc$.  We also suggest that
it is not unreasonable that the disc was initially somewhat larger
than the tidal radius when the SMC started to experience tidal
stripping.

We performed stability tests on all initial SMC models, in the absence
of any external potential (the equilibrium run), to ensure the initial
models were indeed in equilibrium.  The SMC models were evolved using
the \GCD{} parallel tree N-body code described by \citet{kg03a} for
$2.5 \Gyr$ (the interaction run).  We encountered only a minimal
degree of disc heating and newly-introduced spiral structure
(generally with two symmetric arms).

\begin{table*}
  \caption[Parameters for the SMC]{Empirical and simulated
    characteristics of the SMC.}
  \begin{minipage}[P]{18cm}
    \begin{tabular}{lllll}
      Parameter & Fiducial model & Surveyed range & Observation &
      References \\
      \hline
      \HI{} Mass within SMC ($\msunns$) & -- & -- &  $0.35$ -- $0.56\e9$ & (1),(2),(3),(4)\\ 
      \HI{} Mass within MS ($\msunns$) &  -- & -- & $0.1$ -- $0.5\e9$ & (5),(6)\\
      \HI{} Mass within LAF ($\msunns$) &  -- & -- & $0.03$ -- $0.06\e9$
      & (7),(8)\\
      \HI{} Mass within ICR ($\msunns$) &  -- & -- &  $0.05$ -- $0.65\e9$ & (9),(10),(11) \\
      Initial \HI{} mass ($\msunns$) & $1.5\e9$ & $1.1$ -- $2.0\e9$ & 
      $0.5-1.8\e9$\footnote{The sum of the current observed \HI{} masses
        within the SMC, MS, LAF and ICR.} &  \\
      Stellar Mass within SMC ($\msunns$) & -- & -- & $0.58$ -- $1.8\e9$
      &  (12),(13) \\
      Total Mass within SMC ($\msunns$) & -- & -- & $0.9$ -- $2.4\e9$ &
      (14),(15),(16),(17) \\
      Initial total mass ($\msunns$)&
      $3\e9$ &
      $3.0$ -- $3.6\e9$ &   
      $1.1$ -- $3.6\e9$\footnote{The sum of
        the current total mass within the SMC (obtained dynamically)
        and \HI{} masses of MS, LAF and ICR.  This is likely
        underestimated because some dark matter may have been stripped
        from the SMC, as well.}
      \\
      Initial radius of \HI{} disc ($r_{\rm d,95}$; \kpcns)
      & $7$ &  $4$ -- $7$ &
      $10$ -- $12$\footnote{The observed values for SMC-like dwarfs (see text
        for discussion).\label{ftn:smclikedwarf}} 
      & (18) \\
      Initial scale length of \HI{} disc (\kpcns)
      & $0.5 \times r_{\rm d,95} = 3.5$ & $0.2$ -- $0.5 \times r_{\rm d,95}$ &
      $1.5$ -- $4.5$\footref{ftn:smclikedwarf} & (19) \\
      Initial scale height of \HI{} disc (\kpcns)
      & $0.05 \times r_{\rm d,95} = 0.35$ &
      $0.025$ -- $0.075 \times r_{\rm d,95}$ & -- &  \\
      Initial velocity disp.\ of \HI{} disc\footnote{Central velocity
        dispersion.} (\kmsns)
      & $25$ &  $25$ -- $35$ & $\sim 25$ & (20) \\
      Initial radius of DM halo (\kpcns)
      & $1 \times r_{\rm d,95} = 7$ &
      $1$ -- $1.3 \times r_{\rm d,95}$ & -- & \\
    \end{tabular}
  \end{minipage}

  \begin{flushleft}
    References: (1)~\citet{bks+05}, (2)~\citet{ssd+99}, (3)
    \citet{ssj04} who includes He mass, (4)~\citet{psf+03}, 
    (5)~\citet{psf+03}, (6)~\citet{bks+05}, 
    (7)~\citet{bks+05}, (8)~Upper limit is obtained by summing LAF ($b<0\degr$) mass in
    \citet{pgs+98} with HVC clouds EP and WD (but not WE) complexes in \citet{ww91},
    (9)~\citet{msz+03}, (10)~\citet{psf+03}, (11)~\citet{bks+05}, (12)~\citet{ssj04}, (13)~\citetalias{yn03}, 
    (14)~\citet{dlf+85}, (15)~\citet{hsa89}, (16)~\citet{hcm+97}, (17)~\citet{ssj04},
    (18)~\citet{svv+02}, (19)~\citet{svv+02}, 
    (20)~this paper (\Fig{skymom12})
  \end{flushleft}
  \label{tab:paramranges}
\end{table*}

\subsection{Interaction simulations}
\label{sec:simulation/evolve}

We adopt an SMC-centric non-inertial (but non-rotating) coordinate
system for our interaction simulations, one in which the SMC disc lies
on the $x$~--~$y$ plane.  The orbits of the LMC and MW from
\Sec{simulation/orb} are translated and rotated to this coordinate
system, and the SMC model constructed in \Sec{simulation/smc} evolved
within this new coordinate system under the influence of the now
``orbiting'' MW and LMC.  There are two degrees of freedom for the
current inclination angle of the SMC, both currently unknown.  We
define $\theta$ and $\phi$ following fig.~1 of \citetalias{gn96},
survey the full range of both, and and find that the final features of
the simulated MS are sensitive to this angle.  We adopt in the
fiducial model an angle of $(\theta, \phi) = (45\degr, 210\degr)$, as
discussed in \Sec{paramdep}.  This choice leads to a trailing tidal
stream with an orientation consistent with the observed MS, and a
leading arm with shape qualitatively similar to the observed LAF.
This angle is mildly different from that adopted in \citetalias{gn96},
$(\theta, \phi) = (45\degr, 230\degr)$, but we found that the new value
leads to a better match to the HIPASS dataset (data which were not
available to \citetalias{gn96})

The particles in the SMC are assumed to be collisionless and their
dynamical evolution calculated with \GCD{}.  The acceleration applied
to the $i$-th particle is described (\citetalias{gn96}) by
\begin{eqnarray}
  \lefteqn{
    \ddot{{\bf r}_{i}} = -G\sum^n_{j \ne i}
    \frac{m_j\left({\bf r}_i - {\bf r}_j\right)}
    {\left(\left|{\bf r}_i - {\bf r}_j\right|^2 +
        \epsilon^2\right)^{3/2}}} \nonumber \\
  & \mbox{} + {\bf F}_{\rm MW}
  \left({\bf r}_i - {\bf r}_{\rm MW}\right) +
  {\bf F}_{\rm LMC}
  \left({\bf r}_i - {\bf r}_{\rm LMC}\right) \nonumber \\
  & \mbox{} -
  {\bf F}_{\rm MW}\left(-{\bf r}_{\rm MW}\right) -
  {\bf F}_{\rm LMC}\left(-{\bf r}_{\rm LMC}\right)
  ,
\label{eqn:particleforce}
\end{eqnarray}
where $\epsilon$ is the softening length and $m_{j}$ is the mass of
the $j$-th particle.  The position of ${\bf r}$ is measured in the
SMC-centric coordinate frame.  The first term is the self-gravity of
the SMC particles; the second and third terms are the forces on the
particle resulting from the Galaxy and LMC, and can be derived from
the respective potentials in \eqnsII{potMW}{potL,SMC}:
\begin{equation}
  {\bf F}_{\rm MW}({\bf r})=-\frac{V_{\rm c}^2}{|{\bf r}|^2}{\bf r},
  \label{eqn:forceMW}
\end{equation}
\begin{equation}
  {\bf F}_{\rm LMC}({\bf r})=-\frac{G M_{\rm L}{\bf r}}{\left({\bf r}^2 +
      K_{\rm L}^2\right)^{3/2}} .
  \label{eqn:forceLMC}
\end{equation}
The final two terms arise from needing to correct for the integration
of the equations of motion in a non-inertial reference frame centred
on the SMC.

We use 200,000 disc and 200,000 halo particles to describe the SMC.
This corresponds to a resolution $\sim 30$ times greater than that
employed by \citetalias{gn96}, \citetalias{g99} and \citetalias{yn03}.
Such high-resolution allows us to examine features of the MS, LAF, and
SMC, in a manner not previously possible, since smaller fractional
differences in particle density become statistically significant.  We
adopt approximately equal masses for the halo and disc particles of
$7.6\e3\msun$ and $7.4\e3\msun$ respectively, and employ softening
lengths $\epsilon_{d,h}$ of $\sim 65 \pc$.

\begin{figure*}
  \begin{center}
    \begin{tabular}{cc}
      \includegraphics[width=8cm]{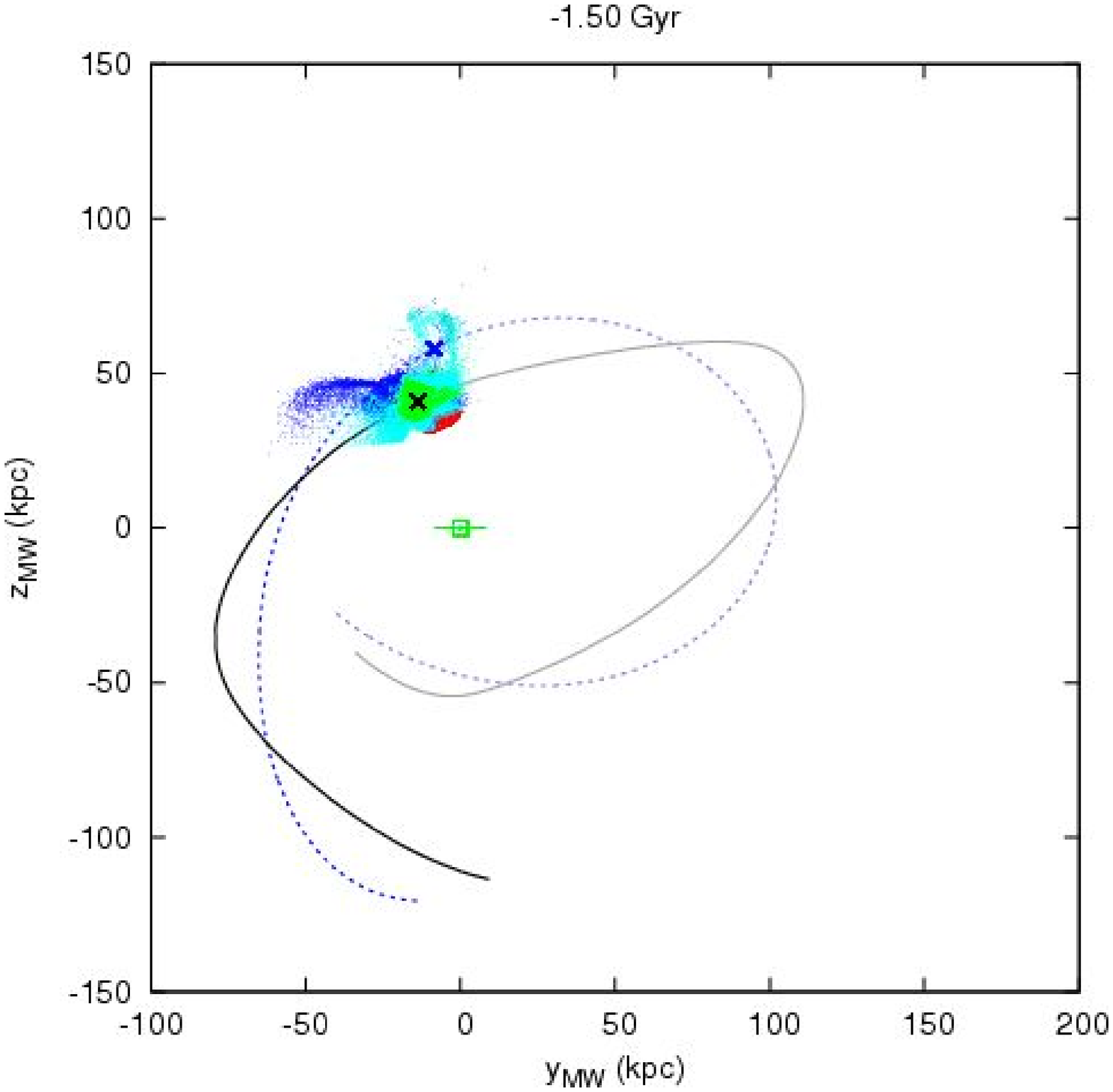} &
      \includegraphics[width=8cm]{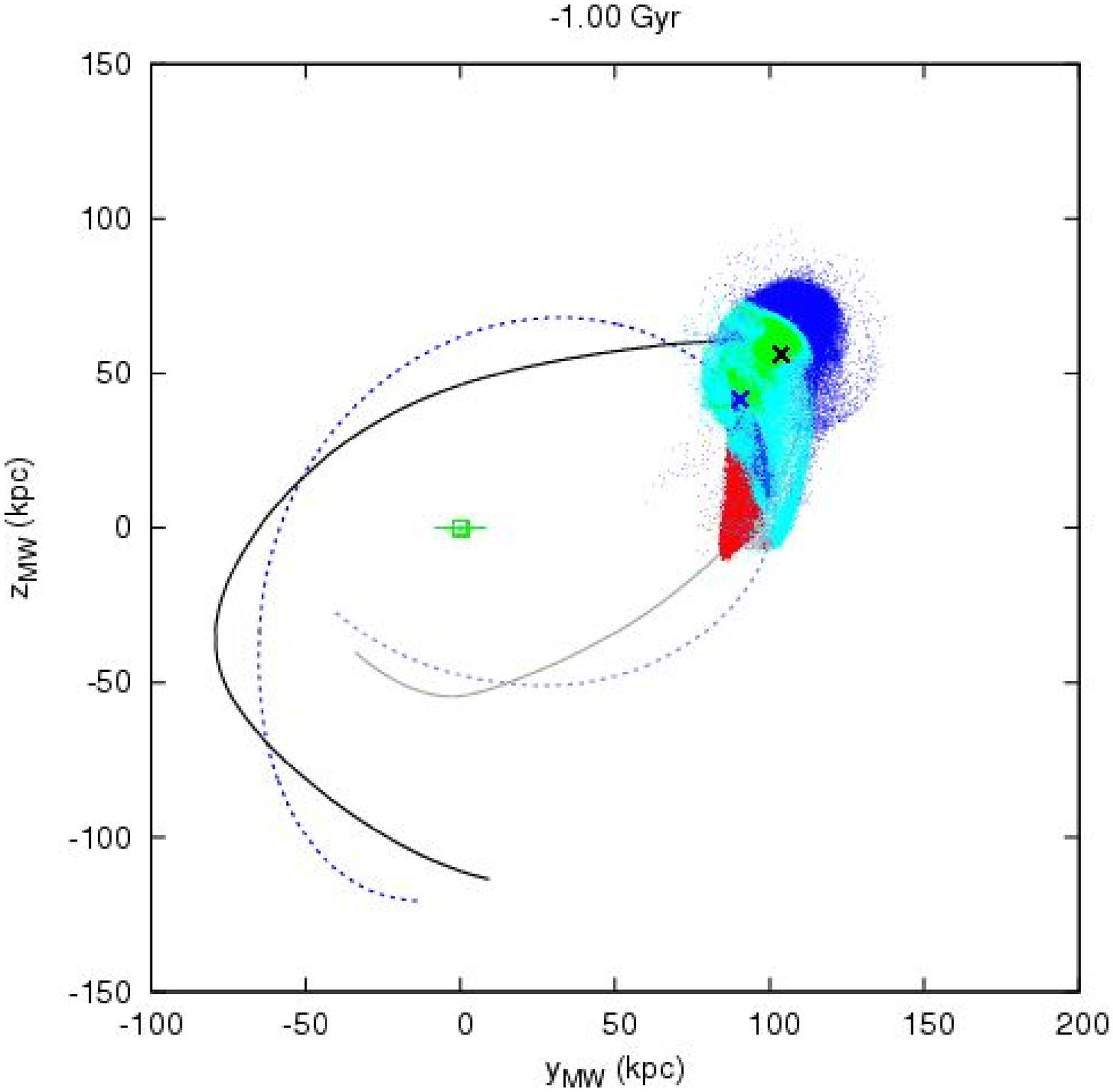} \\
      \includegraphics[width=8cm]{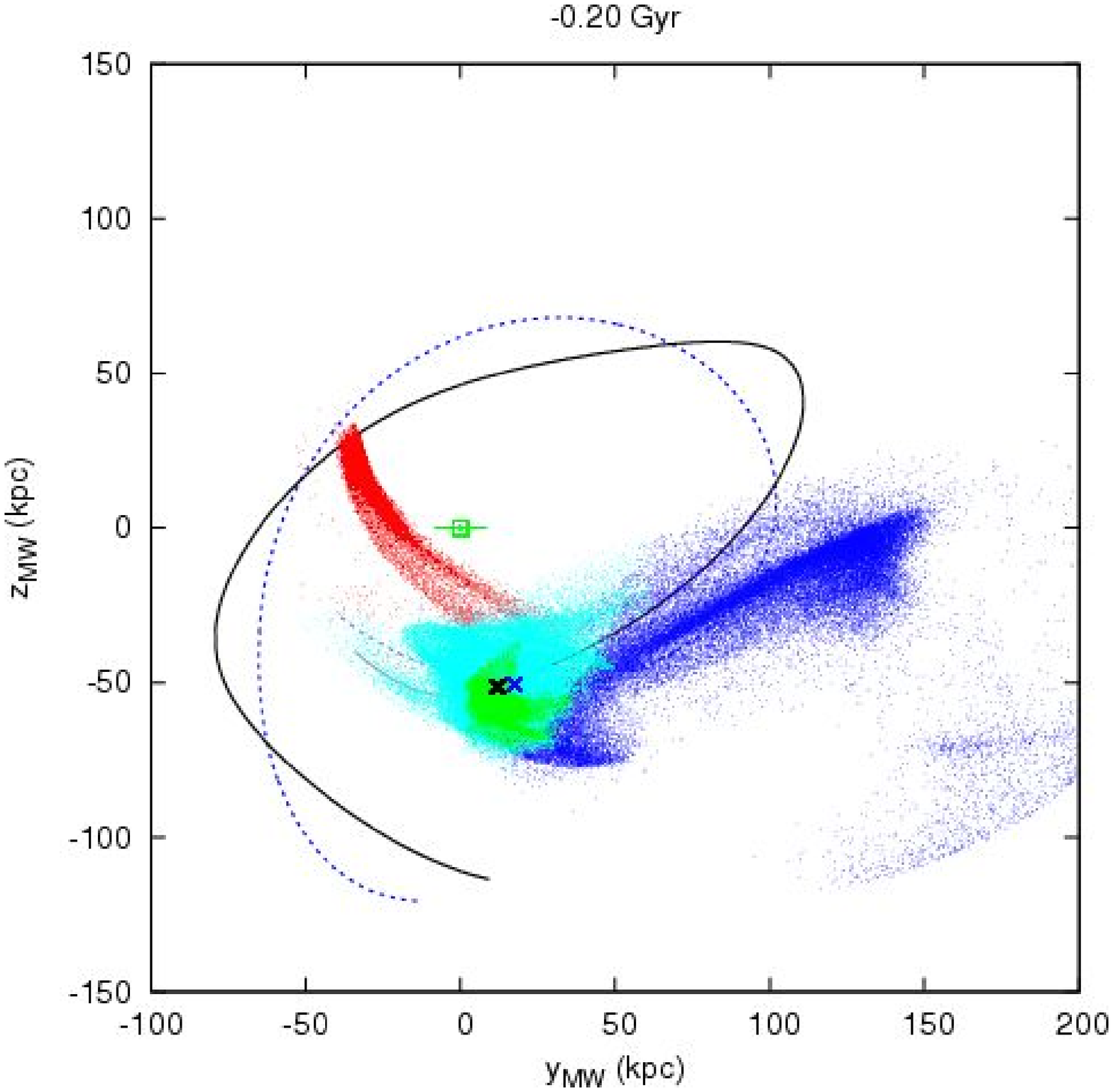} &
      \includegraphics[width=8cm]{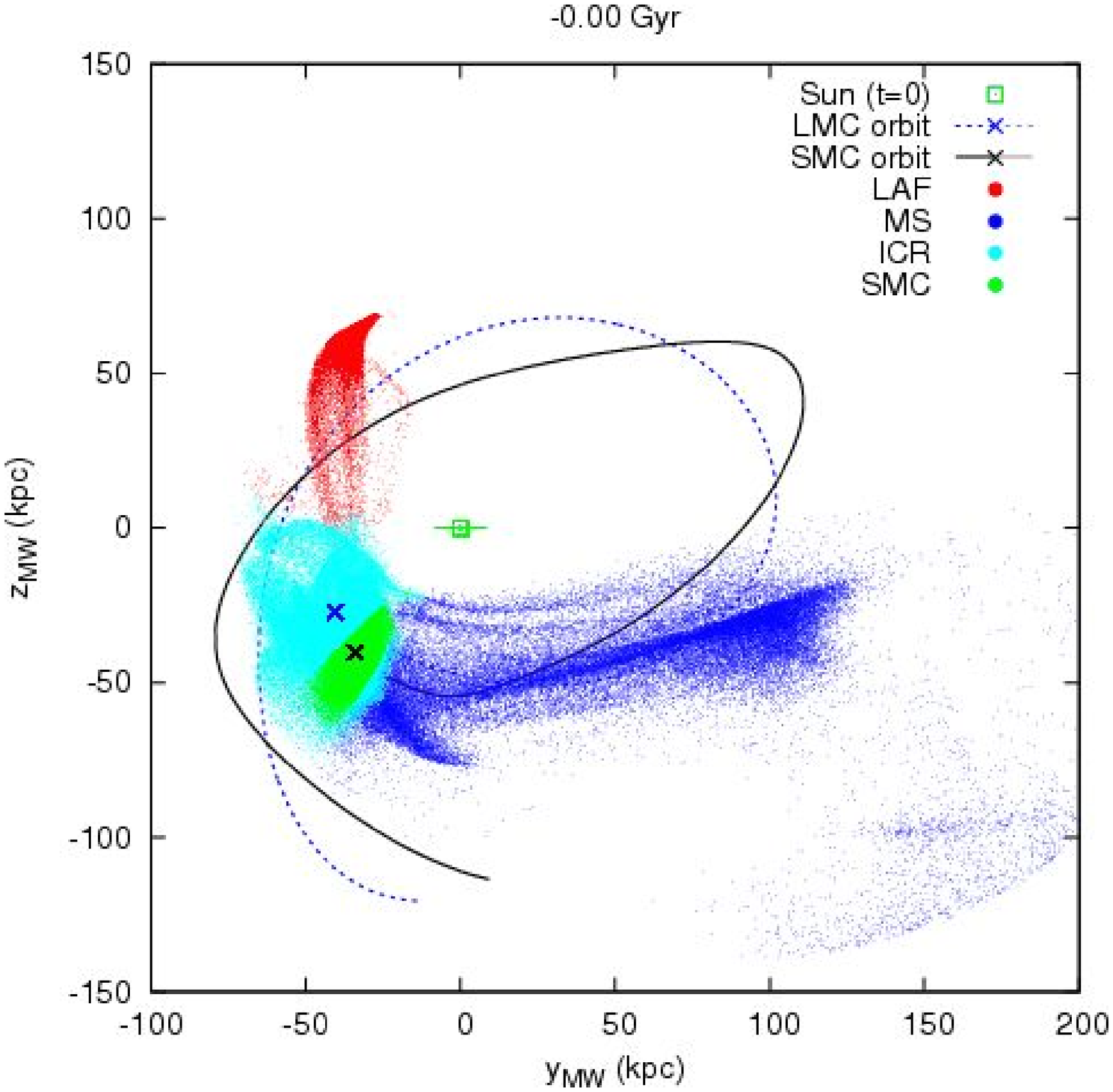} \\
    \end{tabular}
  \end{center}
  \caption[Particle configuration at various timesteps]{The particle
    configuration in the Galactocentric coordinate system at different
    timesteps.  Different colour dots represent the particles which
    end up in different components at $T=0$ as shown in the
    bottom-right plot.  See the text and \Fig{skymom0} for the
    definitions of the MS, LAF, ICR, and SMC.}
  \label{fig:xyz_smc_fiducial} 
\end{figure*}

\begin{figure*}
  \begin{center}

    \includegraphics[width=8cm]{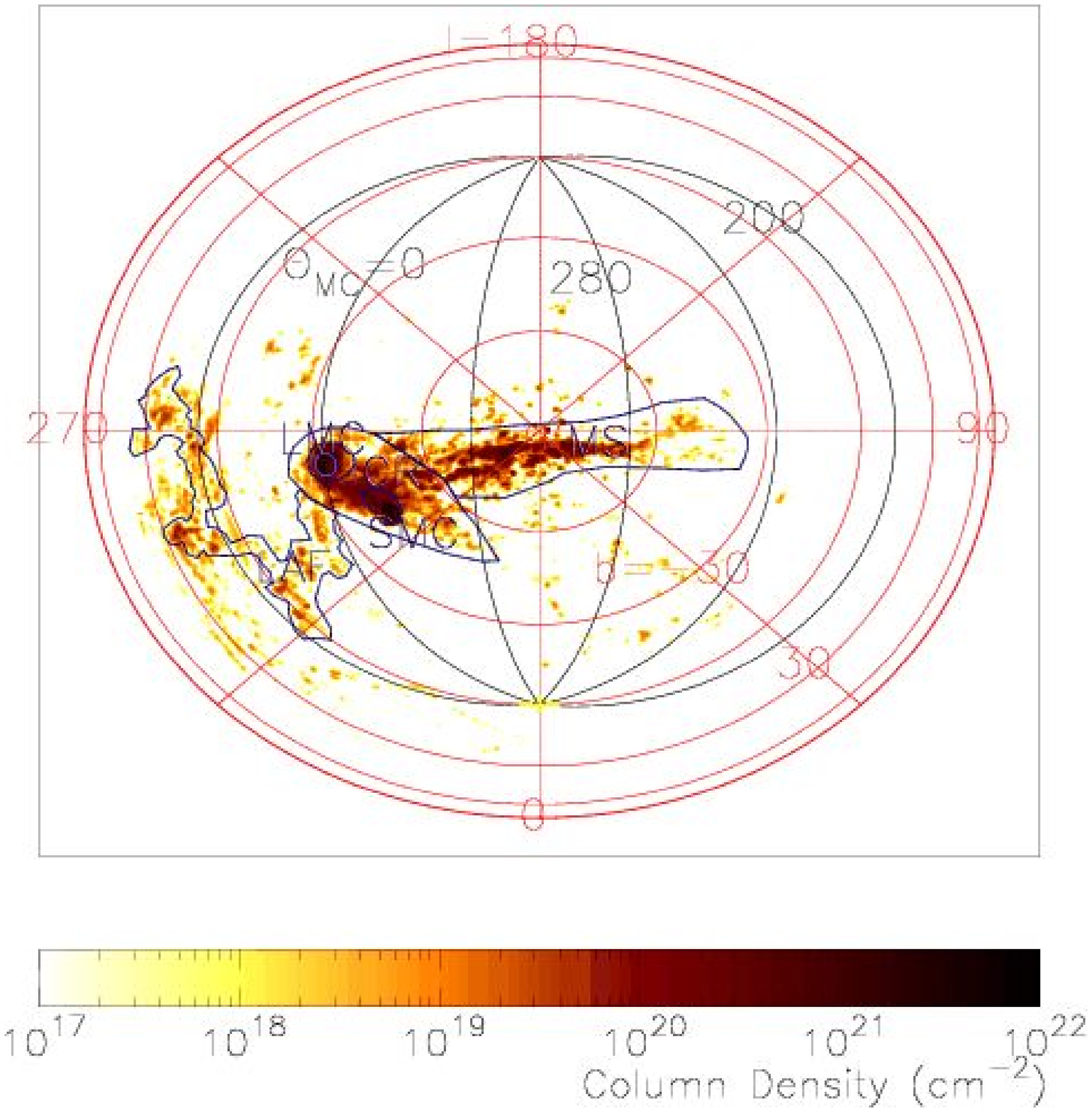}
    \includegraphics[width=8cm]{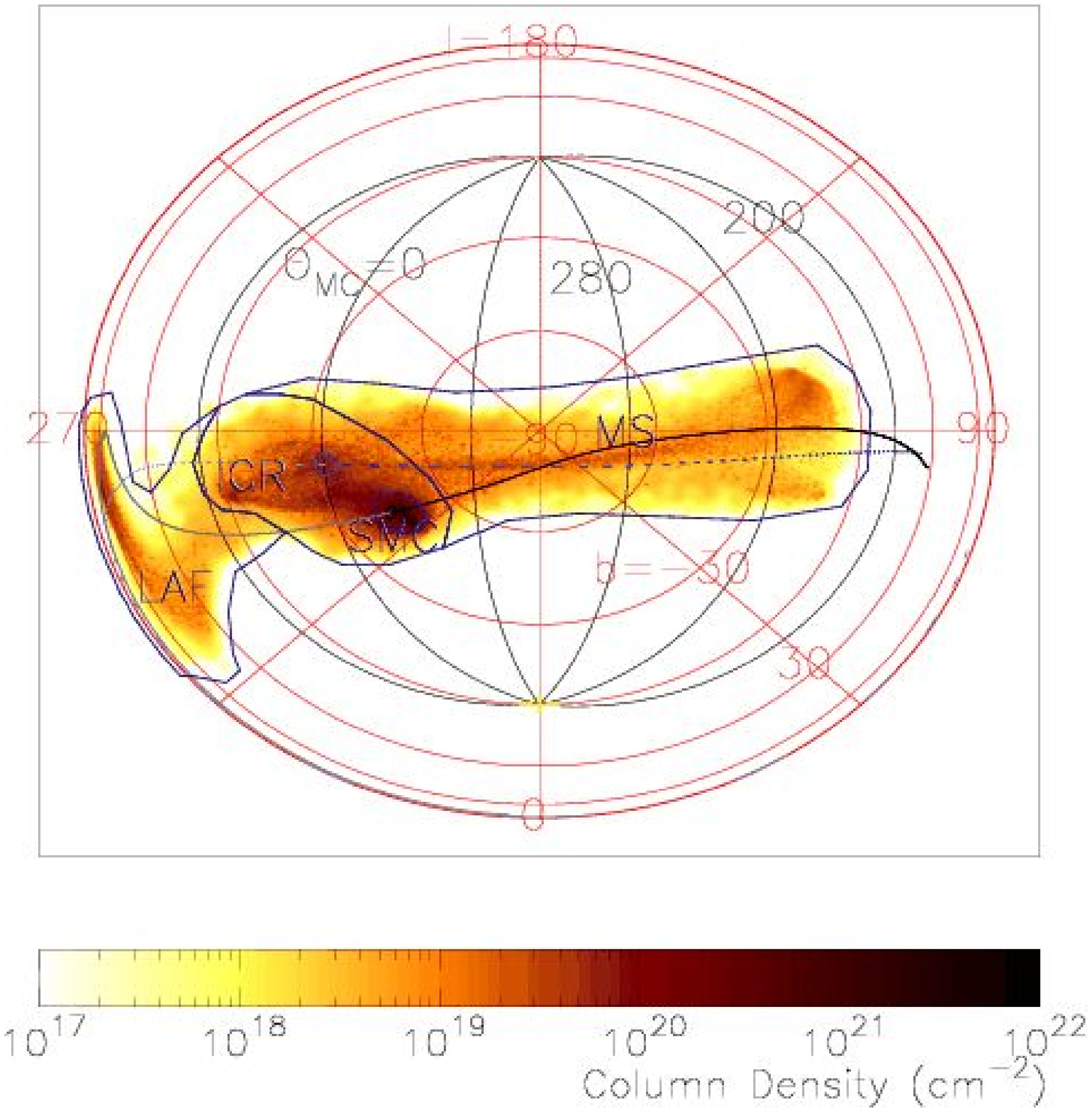}
  \end{center}
  \caption[\HI{} column density map]{\HI{} column density map (i.e.\
    zeroth moment map) of the empirical data (left) and our fiducial
    model (right).  A full sky Zenith Equal Area projection centred on
    the South Galactic Pole is applied, to preserve flux.  Blue lines
    delineate regions used to quantify masses in
    \Sec{fiducial/HIcolumndens}.  In the right panel, the enclosed
    regions with labels of MS, LAF, ICR, and SMC are arbitrarily
    chosen, and are used in \Fig{xyz_smc_fiducial}.  The current
    positions of the SMC and LMC are represented by black and blue
    ellipses, and in the right panel, the past and future $1\Gyr$
    histories of the SMC and LMC orbit are denoted by black (solid)
    and blue (dashed) lines, respectively (the future $1 \Gyr$ is
    shaded in a lighter colour).  Lines of constant Magellanic
    Longitude are drawn in black, with the LMC lying on the $0\degr$
    meridian.}
  \label{fig:skymom0} 
\end{figure*}

\section{Fiducial Model}
\label{sec:fiducial}

In this section, we show the results of our fiducial model, the basic
parameters for which are listed in \Sec{simulation} and
\Tab{paramranges}.  Since we are primarily interested in the formation
of the MS and LAF, which we assume are both stripped \HI{} gas
components from the SMC disc, our analysis focuses on the particles
which were initially within the original SMC disc.  In what follows,
our simulation products are compared closely with the empirical HIPASS
dataset.

\subsection{Evolution}
\label{sec:fiducial/evolution}

In \Fig{xyz_smc_fiducial}, we present a series of snapshots in the
Galactocentric coordinate system, with the orbit of the MCs overlayed.
Consistent with earlier models (e.g.\ \citetalias{gn96};
\citetalias{yn03}), when the SMC experienced a close encounter with
the MW and the SMC $\sim 1.5 \Gyr$ ago, the edge of the disc of the
SMC began to be drawn out, which formed the tidal features that later
became the LAF and MS.  By $T = -1.0 \Gyr$, at the subsequent
apo-Galacticon, the LAF becomes more prominent, whilst the MS was
still under development.  By $T = -0.3 \Gyr$, much of the initially
stripped material in the leading tidal arm had been pulled back into
the inter-cloud region (ICR), and the material still in the LAF was
brought within $3 \kpc$ of the solar circle.  By this time, the MS and
LAF morphology resemble that seen today.  The next encounter with the
MW and SMC at $T = -0.2 \Gyr$ caused little obvious consequence to
either the MS or LAF.  It did however, cause the dispersion of much of
the material that had been within the SMC.  Much of this material
either ended up in the ICR (although most of the ICR material was
already in an ICR structure before this event), or contributed to the
large spread in radial extent of the SMC.  At the current time, the
ICR extends radially from $\sim 30$ to $\sim 80 \kpc$, and the SMC
from $\sim 45$ to $\sim 60 \kpc$.

We also ran the simulation ``forward'' in time to $T=+0.25\Gyr$ and
found that the ICR undergoes mass loss with material being dragged out
into two tidal tails, separated radially and kinematically from the
main MS and LAF features.  This is consistent with what is expected by
\citet{bks+05}.  By $T=+0.25\Gyr$, the material in the ICR has been
dragged $55\kpc$ towards the MS in the plane of the sky, giving an
angular separation of $35\degr$ between the SMC and tip of the ICR.

\subsection{\HI{} column density distribution}
\label{sec:fiducial/HIcolumndens}

\begin{figure*}
  \begin{center}
    \includegraphics[width=8cm]{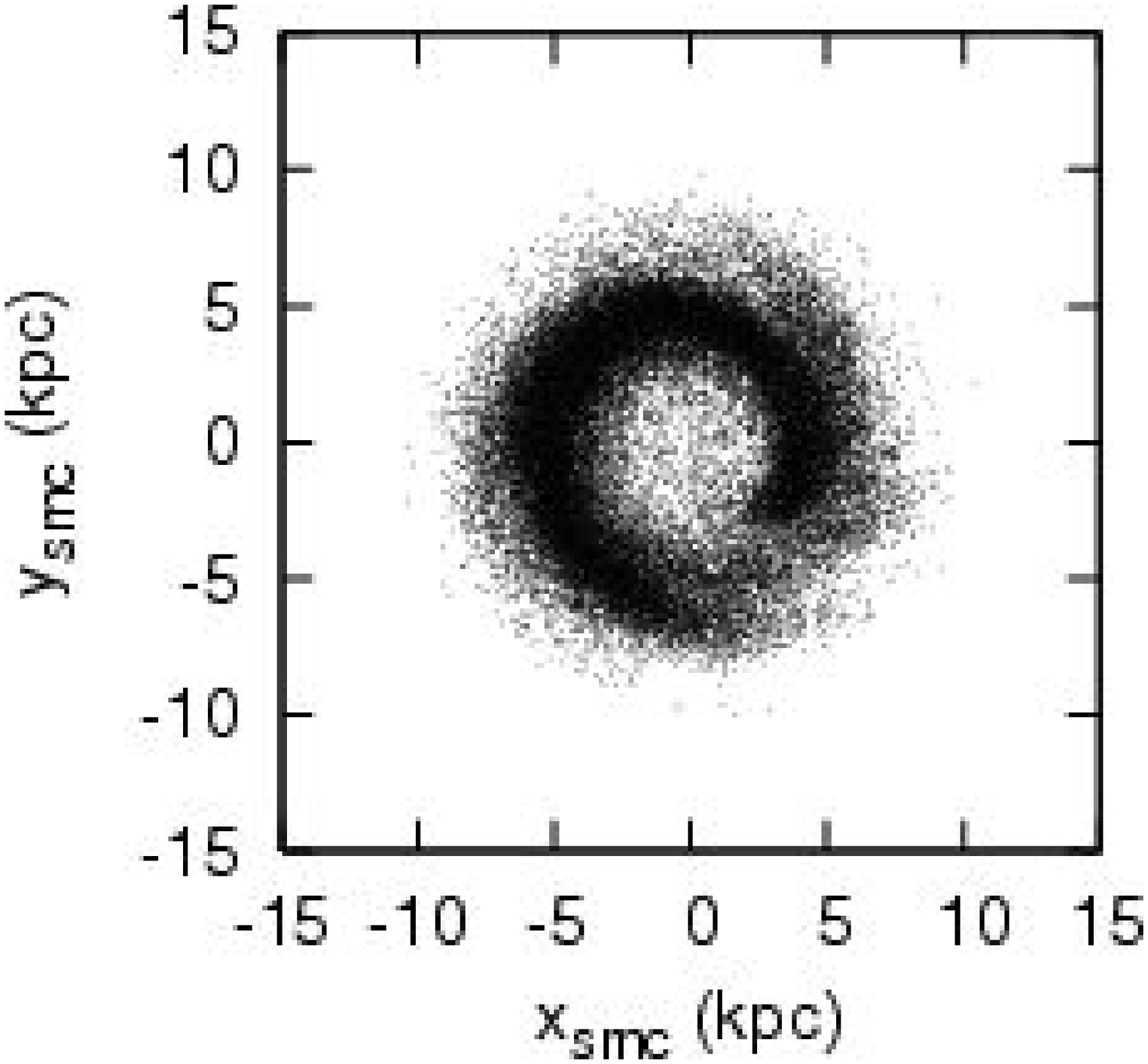}
    \includegraphics[width=8cm]{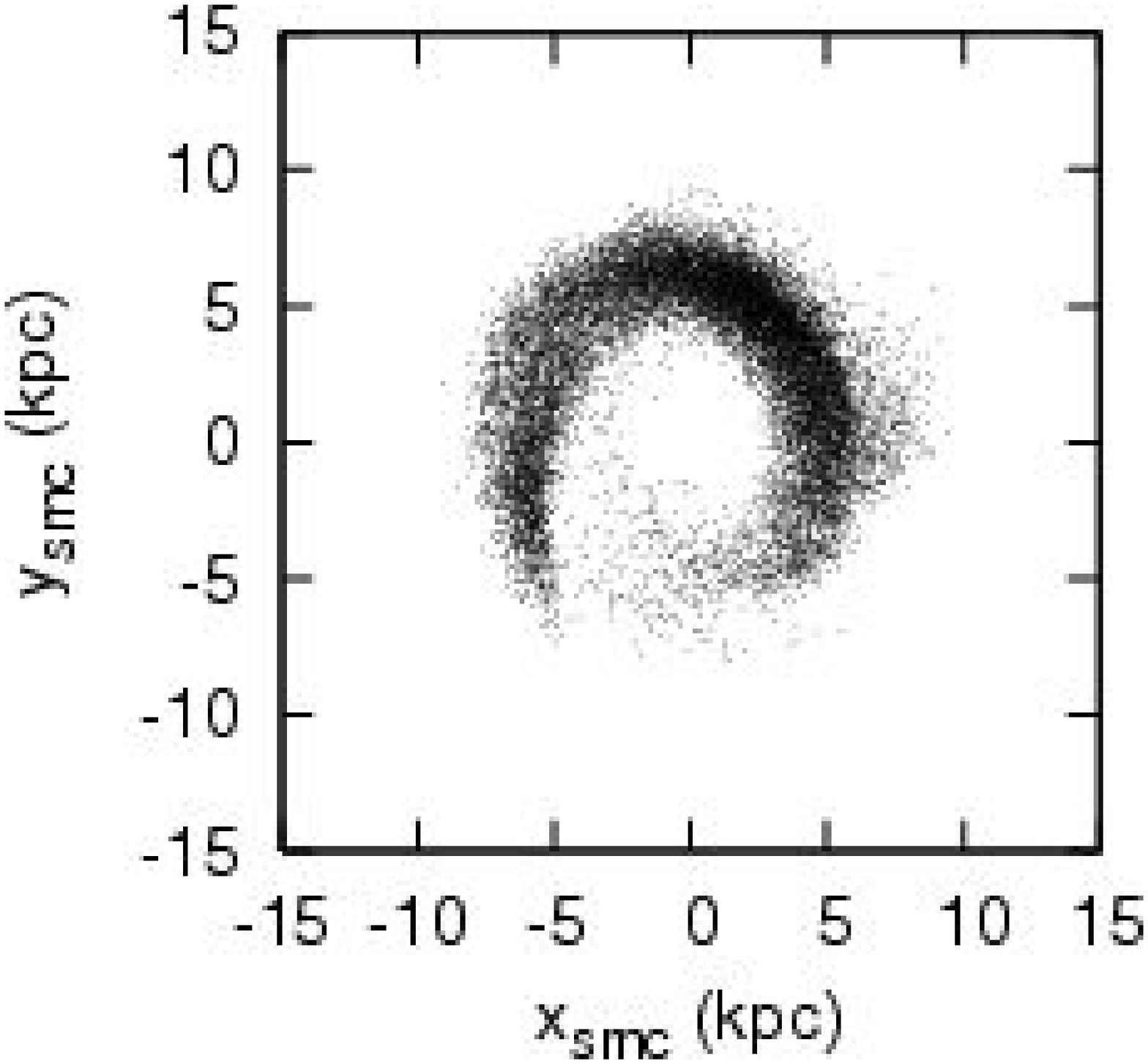}
  \end{center}
  \caption[Initial distribution of particles that reside in MS and LAF
  in final timestep]{The distribution of the particles which end up in
    the MS (left) and LAF (right) at $T=0$ in the face-on view of the
    initial ($T=-2.5 \Gyr$) SMC disc.  For the definition of the MS
    and LAF see \Fig{skymom0}.}
  \label{fig:iconfig} 
\end{figure*}

In this section, we compare the present-day \HI{} column density
distribution between the empirical HIPASS dataset \citep{pgs+98} and
our simulation results.  \Fig{skymom0} displays the \HI{} column
density map on a Zenith Equal Area (ZEA) projection.  Here, we
arbitrarily define the regions corresponding to the MS, LAF, ICR, and
SMC as shown in the right panel of \Fig{skymom0}.  We converted the
observed 21cm flux to \HI{} column density, after \citet{bsd+01}.  To
construct \HI{} column density maps from the simulation results, we
assume that the disc particles in the SMC are purely gaseous, and the
\HI{} mass fraction is 0.76.  The column densities within the SMC
and ICR region will be somewhat overestimated, as we neglect currently
any associated stellar and ionised components.  On the other hand,
\Fig{iconfig} demonstrates that most particles in the MS and the LAF
at $T=0$ originate in the outer edge of the initial SMC disc, where
there was likely a lack of stars.  To date, searches for a putative
stellar component to the MS have proven unsuccessful (e.g.\
\citealp{ap76a,ap76b,er82,bh83}).  Regardless, the \HI{} fraction of
0.76 remains technically an upper limit, and thus our predictions
should considered as upper limits.  For these reasons the comparisons
between the simulation and the HIPASS data focus mainly upon the
properties of the MS and LAF.

\Fig{skymom0} shows that in our fiducial model the gross features of
the observed MS are reproduced, and the LAF appears as a consequence
of tidal interactions.  This confirms previous studies, such as
\citetalias{gn96} and \citetalias{yn03}, which suggested that the MS
and LAF features originate from gas stripped from the SMC disc by the
tidal interaction with the LMC and MW.

The left panel of \Fig{closeuplaf}, itself a new representation of the
HIPASS dataset, demonstrates that the observed leading arm extends
above the Galactic plane to latitude $b \sim 30 \degr$.  While the full
extent of the LAF above the plane remains a matter of debate (e.g.\ 
\citealp{bks+05}), the metallicity of its gas is consistent with an
SMC origin \citep{lss+98,ggp+00}, supporting the tidal disruption
scenario for the MS (and LAF).  Furthermore, the observed \HI{} cloud
distribution seems to show ``a kink'' near $(l,b)=(310\degr,0\degr)$,
where two further components of the LAF are seen north of the Galactic
plane (labelled LAF~II and III in \citealp{bks+05}).  Although the
exact position of the kink is inconsistent with the empirical data,
our simulation does naturally predict its existence.

The tail of the observed MS shows spatial bifurcation near
$(l,b)=(300\degr,-70\degr)$ and $(l,b)=(80\degr,-55\degr)$ in
\Fig{skymom0}, with the two components forming an apparent twisting
double helix-like structure \citep{psf+03}.  This bifurcation is not
apparent in previous studies such as \citetalias{gn96} and
\citetalias{yn03}; higher resolution simulations enable us to study
such subtle features.  Since this might be further evidence of the
tidal interaction between the LMC and SMC, we return to this issue
later.

\begin{figure}
  \begin{center}
    \includegraphics[width=8cm]{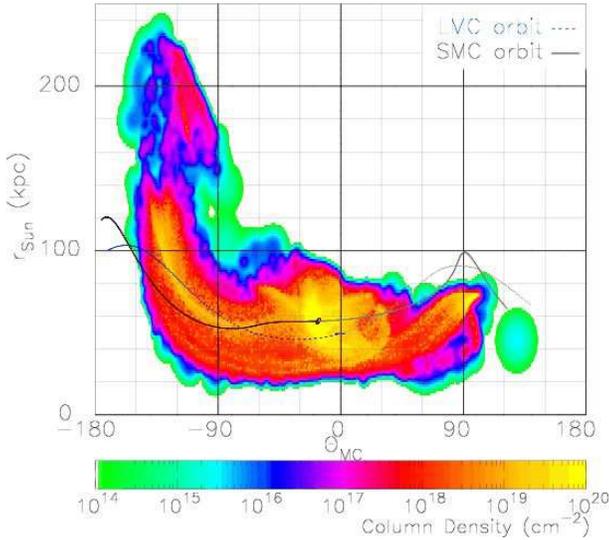}
  \end{center}
  \caption[\HI{} $r_{\rm Sun}$ vs.\ $\lMC$]{The heliocentric
    distance, as a function of the Magellanic longitude, derived from
    our fiducial model.  The black (solid) and blue (dashed) lines
    show the past and future $1 \Gyr$ histories (the future $1 \Gyr$
    is shaded a lighter colour) of distance and Magellanic longitude
    for the SMC and LMC, respectively.}
  \label{fig:rtheta} 
\end{figure}

An advantage of the present work is that the \HI{} data available to
constrain the models are significantly improved beyond that of
\citetalias{gn96} or \citetalias{yn03}.  As in those previous studies,
while the gross features of the observed MS and LAF are reproduced by
our fiducial model, there remain subtle discrepancies between
simulations and data.  While the simulated MS is both broader and more
extended than that observed (and hence the mean column density is
somewhat lower than that encountered in the HIPASS dataset), the
derived \HI{} gas mass (assuming a heliocentric distance of $57 \kpc$)
from HIPASS is within 50 per~cent of that of our simulated fiducial
model ($3.5\e8\msun$ and $2.4\e8\msun$, respectively).  \Fig{rtheta}
shows the heliocentric distance of the simulated MS and LAF against
the Magellanic longitude, $\lMC$ (as defined in \citealp{w01}; lines
of constant $\lMC$ are shown in \Fig{skymom0}).  This demonstrates
that the distance to the simulated MS is not constant, instead
increasing across its length.  Hence negating the distance ambiguity
by obtaining a total flux gives us a fairer comparison than a total
mass.  Doing so results in an observed MS total \HI{} flux of
$4.8\e5\Jykms$, while the fiducial simulated MS has a total \HI{} flux
of $2.3\e5\Jykms$, still within a factor of $\sim 2$ of the HIPASS
data.

Conversely, the simulated LAF has a predicted \HI{} mass and
associated flux ($7.3\e7 \msun$ and $7.9\e4 \Jykms$, respectively)
both factors of $\sim$2 greater than that inferred from the HIPASS
dataset ($3.5\e7\msun$ and $4.6\e4\Jykms$, respectively)\footnote{We
  have defined the LAF and MS regions differently between the
  empirical and simulated datasets, to account for the geometrical
  differences, such as angle, width, and length, between the two.}.
One clear difference between observation and simulation is that of the
geometry of the LAF, in particular that of the projected deflection
angle between the LAF and a Great Circle aligned with the MS proper
(\Fig{closeuplaf}).  In addition, the simulated LAF extends above the
Galactic Plane beyond that observed.  We will discuss possible
solutions to these apparent problems in \Sec{summary}.

\begin{figure*}
  \begin{center}
    \includegraphics[width=8cm]{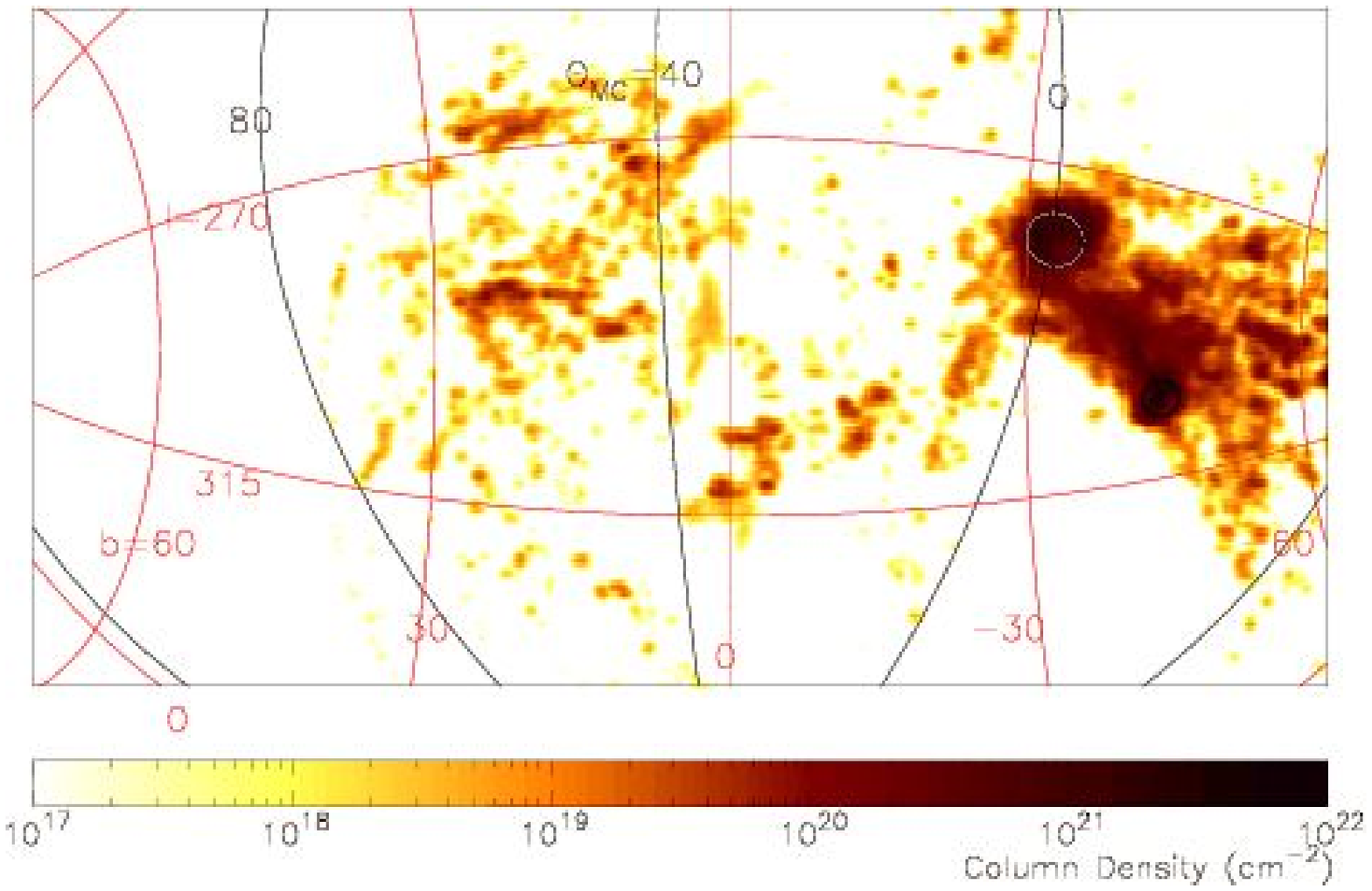}
    \includegraphics[width=8cm]{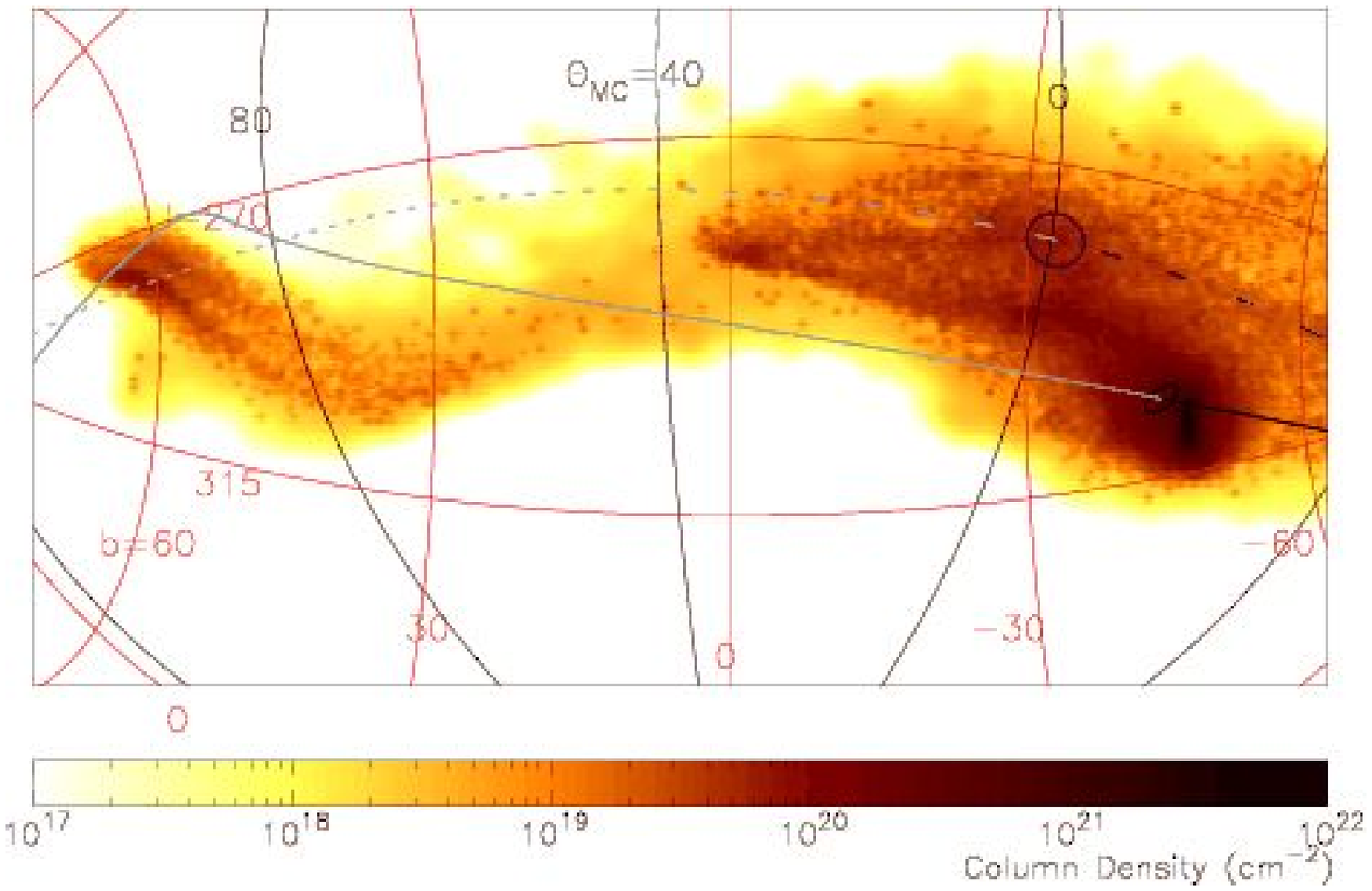}
  \end{center}
  \caption[\HI{} column density map -- closeup view of LAF
  region]{\HI{} column density map of the LAF region derived from the
    HIPASS dataset (left) and that of our fiducial model (right) with
    ZEA projection centred on the LAF.  The future orbits of the MCs
    are denoted by a lighter shading, demonstrating the similarity
    between the orbit of both MCs and the orbit of the LAF.}
  \label{fig:closeuplaf} 
\end{figure*}

\begin{figure*}
  \begin{center}
    \includegraphics[width=8cm]{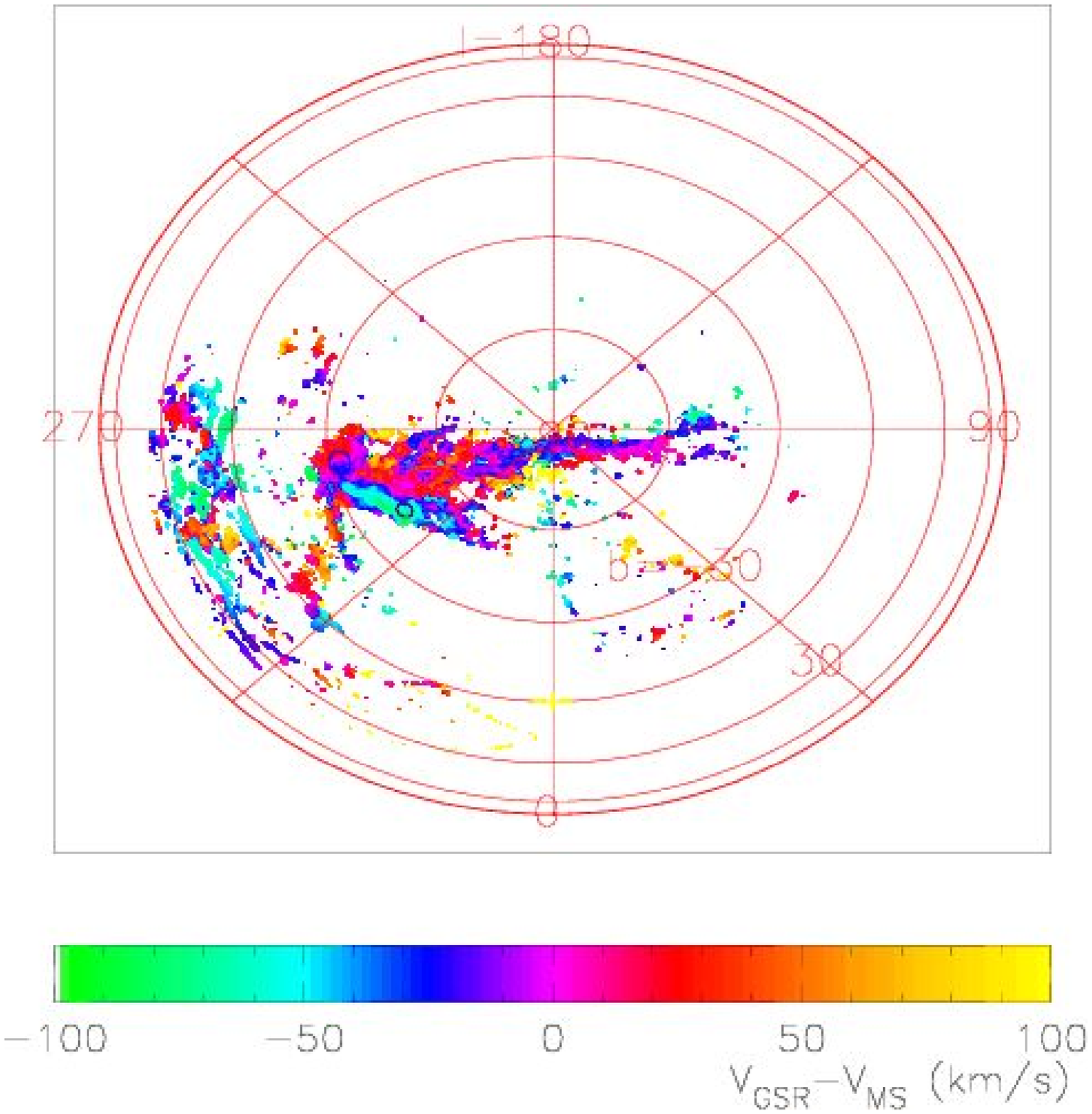}
    \includegraphics[width=8cm]{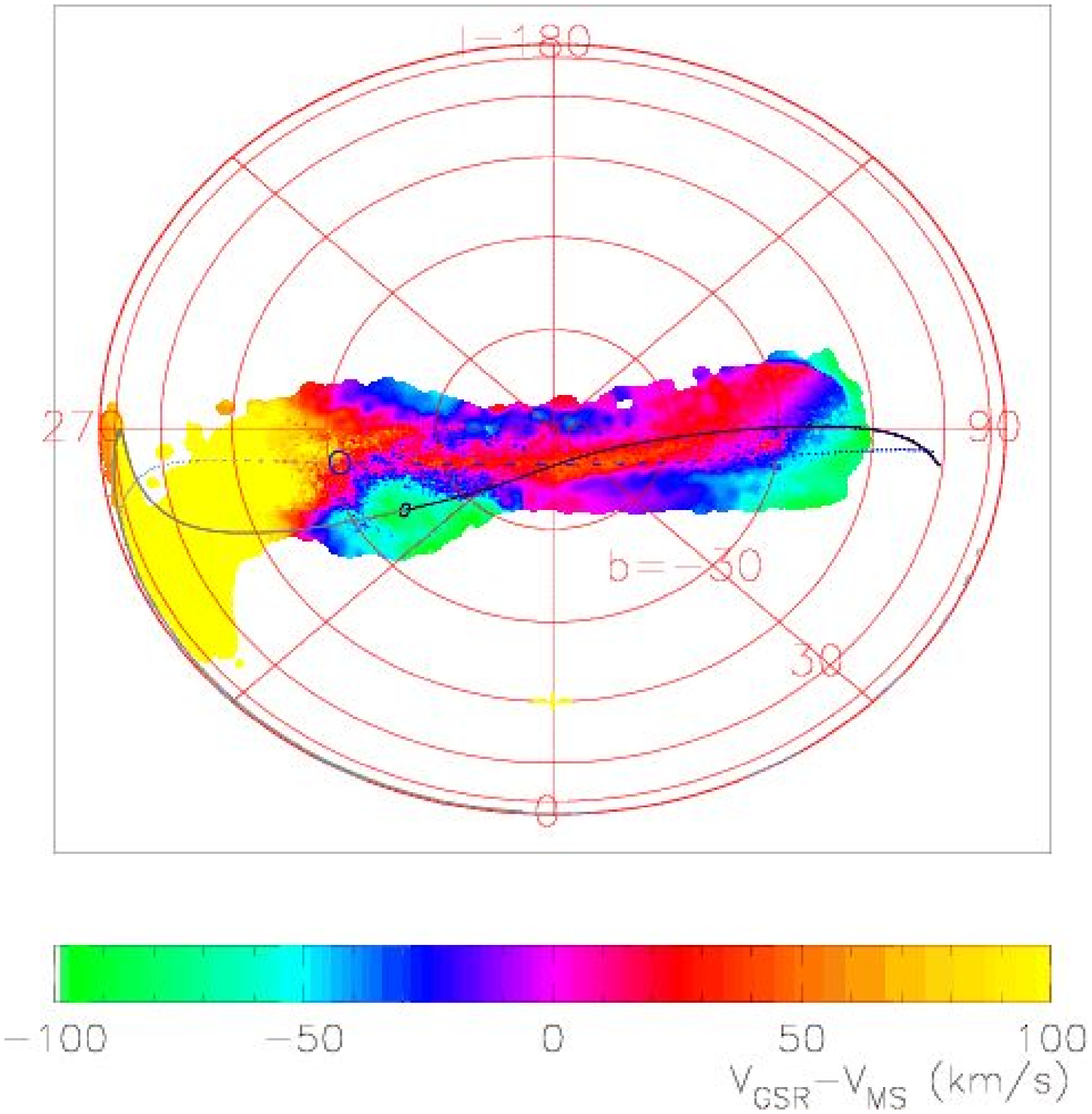} \\
    \includegraphics[width=8cm]{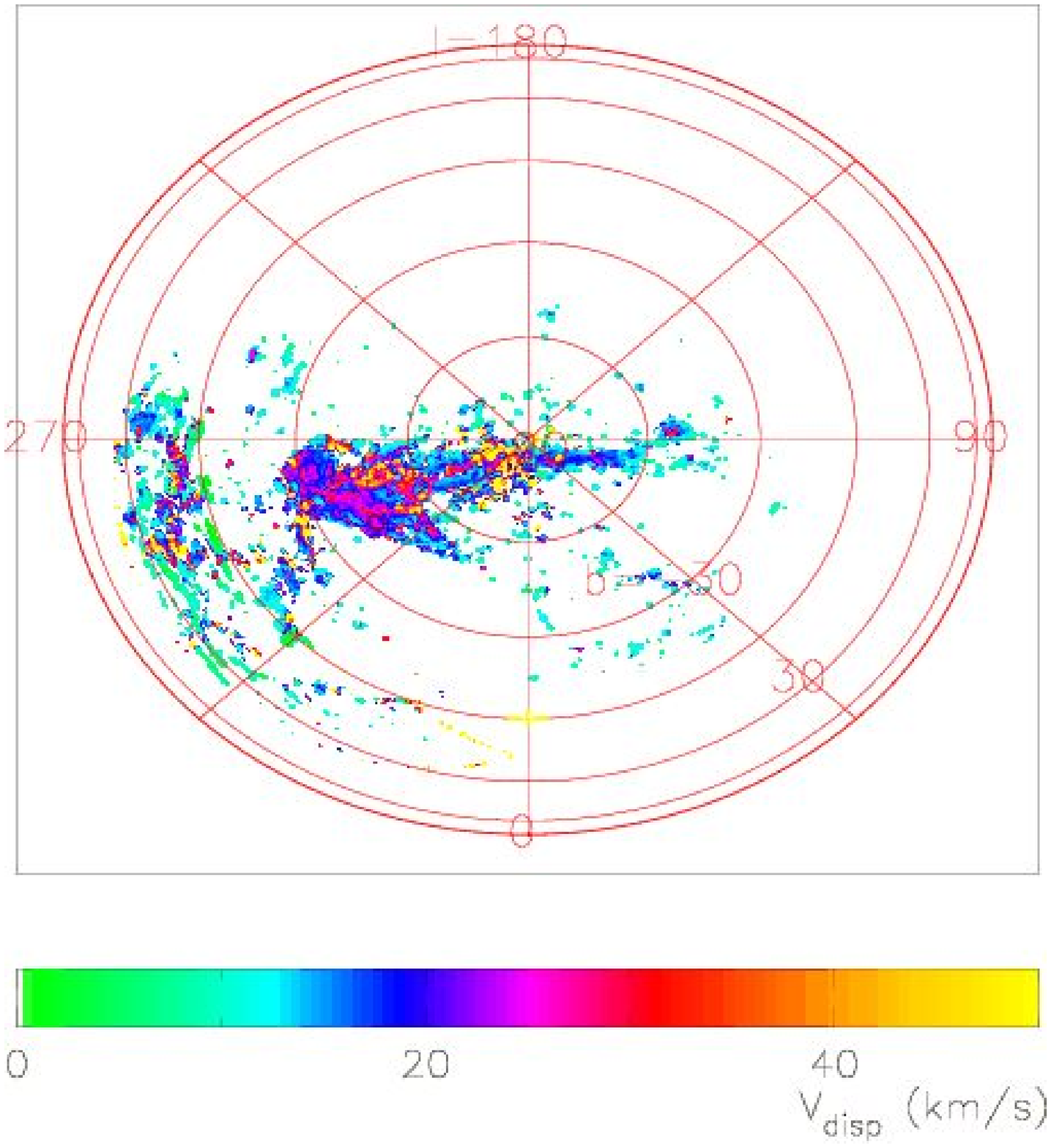}
    \includegraphics[width=8cm]{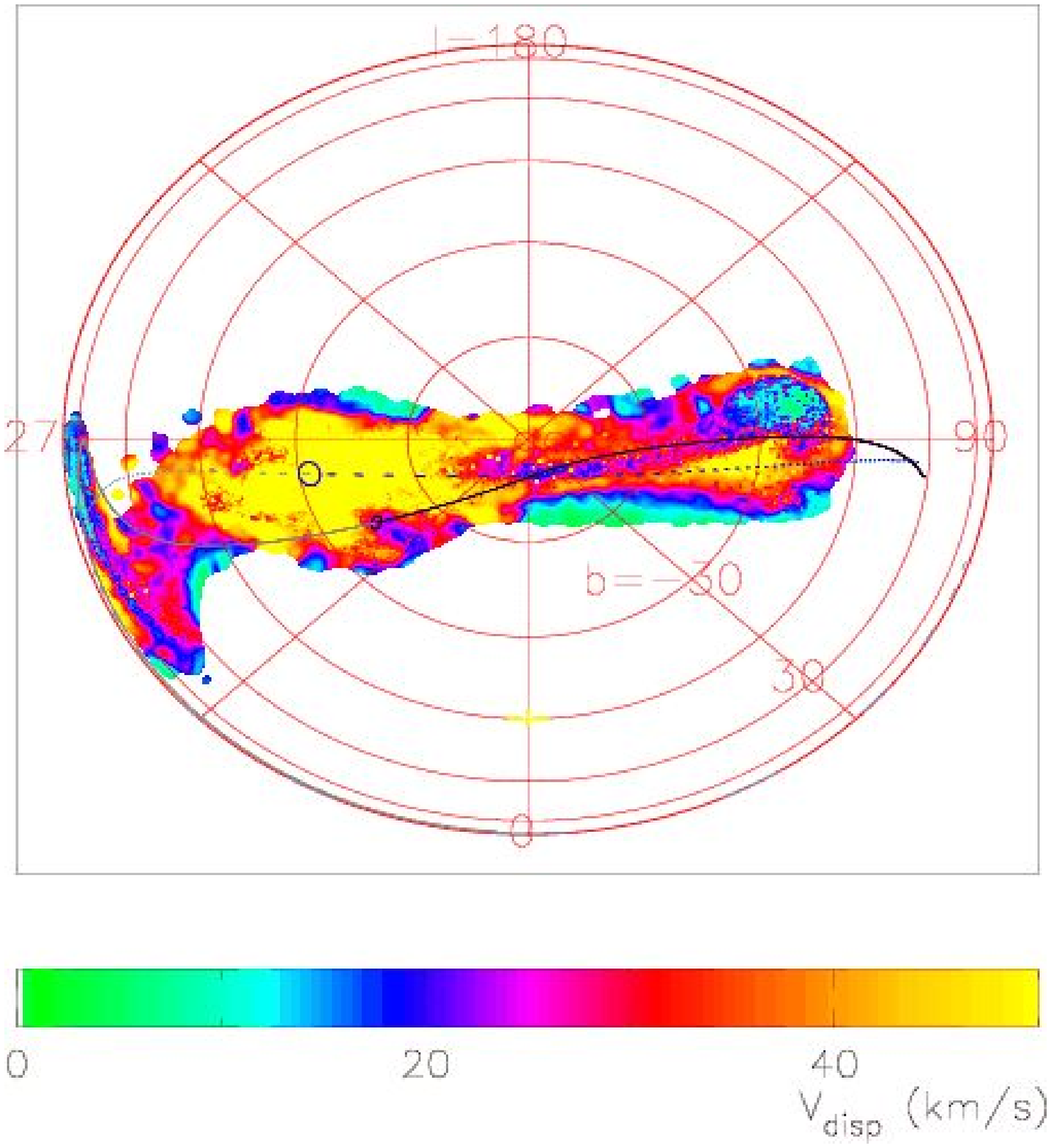}
  \end{center}
  \caption[First and second \HI{} moment maps]{First (upper) and
    second (lower) moment maps derived from the HIPASS dataset (left)
    and that of our fiducial model (right).}
  \label{fig:skymom12} 
\end{figure*}

\begin{figure*}
  \begin{center}
    \includegraphics[width=8cm]{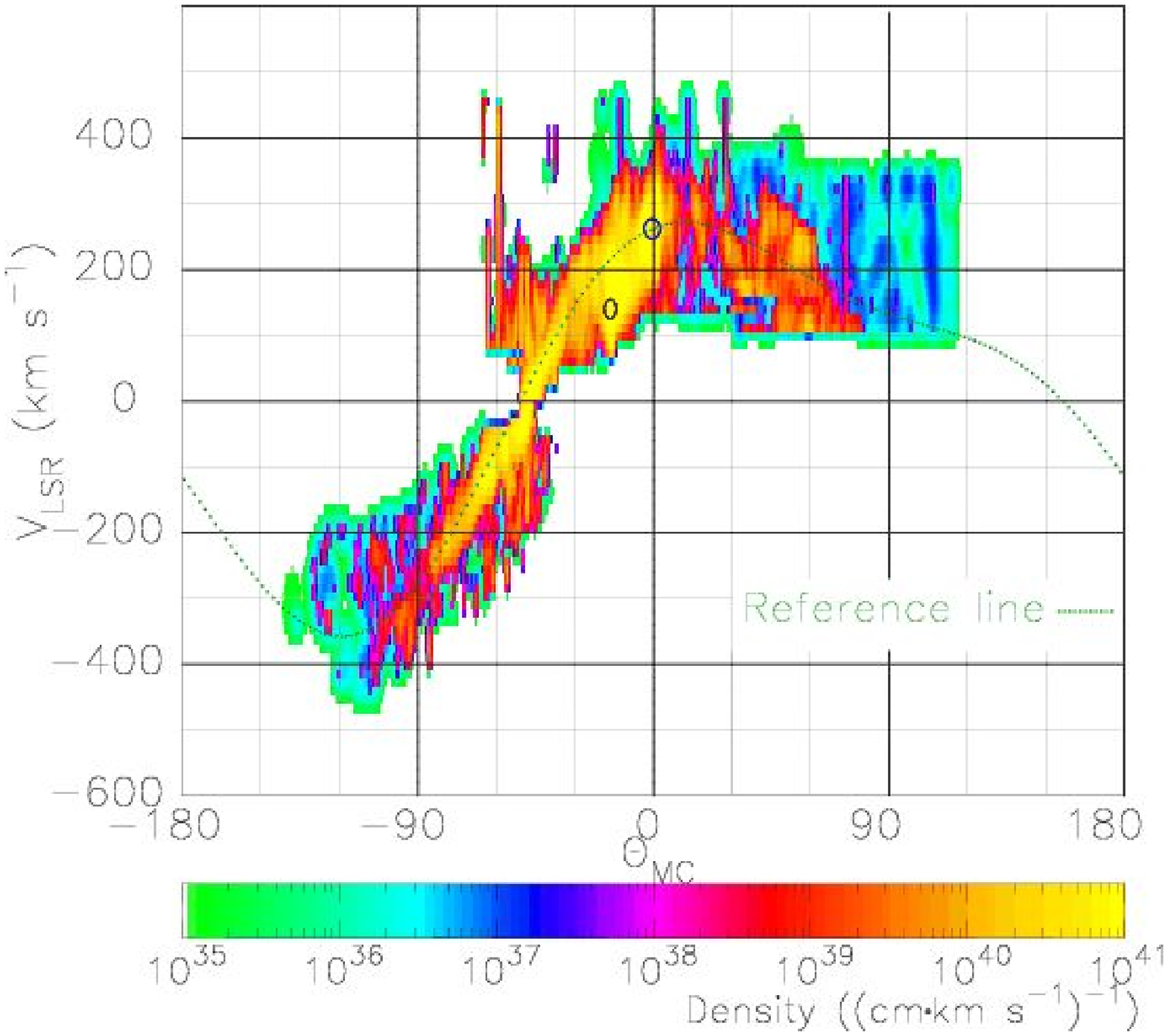}
    \includegraphics[width=8cm]{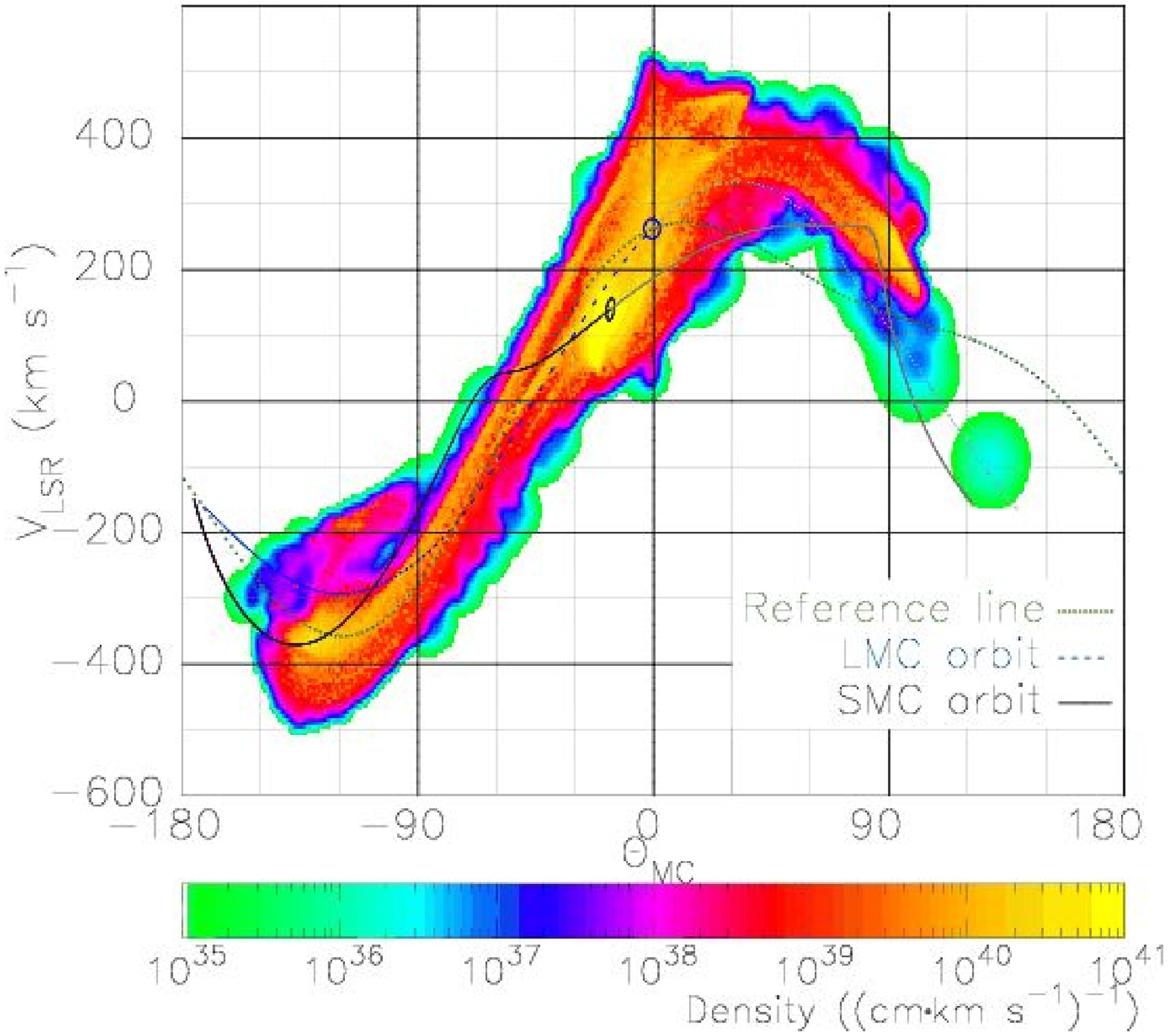}
  \end{center}
  \caption[\HI{} $\vlsr$ vs.\ $\lMC$]{The line-of-sight velocity with
    respect to the Local Standard of Rest $\vlsr$, as a function of
    the Magellanic longitude $\lMC$, derived from the observed HIPASS
    dataset and that of our fiducial model.  The black (solid) and
    blue (dashed) lines show the past and future $1 \Gyr$ histories
    (the future $1 \Gyr$ is shaded a lighter colour) of $\vlsr$ and
    $\lMC$ (where the longitude is defined with respect to the current
    position of the LMC) for the SMC and LMC, respectively.  A
    reference line (dotted green) is drawn to show the velocity
    subtracted equally to yield the top panels of \Fig{skymom12}---see
    text for details.}
  \label{fig:vlsrtheta} 
\end{figure*}

\subsection{\HI{} kinematics}
\label{sec:fiducial/HImoment}

\Fig{skymom12} shows the first and second moment maps for both our
fiducial simulation and that derived from the HIPASS dataset.  In
order to remove the large velocity gradient along the MS (which acts
to obscure fine kinematical details within the maps), the first moment
map is shown as the distribution of the velocity of $\vgsr-\vms$,
where $\vms$ is the mean trend of velocity across the observed MS and
LAF, and is defined in terms of two fitted Fourier components, $\vms =
86 \sin(\lMC+2.5\degr) - 92 \sin(2(\lMC-26.3\degr))$.  The second
moment map is the distribution of the velocity dispersion of \HI{}
gas.

In the first moment map, the observed velocity trend along our
simulated MS is shown to be globally consistent with that of the
empirical data.  \Fig{vlsrtheta} displays $\vlsr$ against the
Magellanic longitude, $\lMC$, and demonstrates perhaps more clearly
that the mean velocity of the simulated MS is consistent with that
observed.  In contrast, the line-of-sight velocity of the simulated
LAF is significantly larger than that observed, although it does
follow the general trend of decreasing velocity at $\lMC \simgt 0$.

The second moment maps seen in the lower panels in \Fig{skymom12}
indicate that the velocity dispersion of the simulated LAF is roughly
consistent with, although slightly higher, than that observed.
However, the velocity dispersion of the simulated MS is somewhat
greater than that inferred from the HIPASS dataset.  This might be due
to the neglect of gas dissipation via radiative cooling in the current
suite of simulations, as will be examined further in Paper~II.

We end by drawing attention to the evidence of a bifurcation in
$\vlsr$ (right panel of \Fig{vlsrtheta}) within the MS and LAF.  The
next section discusses this bifurcation in more detail.

\subsection{Spatial and velocity bifurcation}
\label{sec:fiducial/bif}

Our fiducial model shows a bifurcation in the MS, as seen in the ZEA
spatial distribution of the \HI{} column density of \Fig{skymom0}.
Within the simulation, this bifurcation occurs both radially and
tangentially.  \Fig{skymom0} displays the tangential bifurcation, and
shows that there are two stream components that appear to follow a
twisting topology governed by the orbits of the MCs -- both the orbits
and the bifurcation seem to cross at $(l,b) \sim (45\degr,-80\degr)$.
Radially, the head of the Stream, just behind the MCs, has two
components (\Figs{xyz_smc_fiducial} and \ref{fig:rtheta}).  At the
tail of the MS, there is a third component, well separated from the
other two streams --- extending from $170$ to $220 \kpc$.  It is
visible only through its heliocentric distance from the Sun, and
coincides with the $(l,b)$ position of the tip of the main streams.
Also apparent in \Figs{rtheta} and \ref{fig:closeuplaf}, are the
radial and tangential bifurcations of the LAF, respectively.

The right panel of \Fig{vlsrtheta} demonstrates that the bifurcation
of the MS appears also in $\vlsr$.  Plotted in \Fig{vlsrtheta} are
lines showing the history (and future) of the MC's $\vlsr$.  Only one
of the bifurcated components follows the orbit of the Clouds
(primarily the LMC), while the other possesses a higher velocity.
Interestingly, there is a second velocity component at the position of
the SMC (as well as at the head of the MS and the LAF).  We remind the
reader that \citet{mf84} observed two velocity components within the
SMC itself, with a separation of $\sim 50\kms$.  While the separation
of our two velocity components is much larger, it might indicate that
the two observed velocity components are caused by a similar process.

Snapshots of the simulation, similar to those seen in
\Fig{xyz_smc_fiducial}, hint at the origin of these bifurcations.
Prior to the first major peri-Galacticon at $T = -1.5 \Gyr$, an
encounter with the LMC at $T = -2.2 \Gyr$ drew the particles from the
tip of the SMC disc closest to the LMC (and furthest from the MW),
which resulted in the particles that eventually came to reside in the
most distant MS component, at a Galactocentric distance of $170$ to
$220 \kpc$.  The MS particles did not become ``distinct'' from the SMC
proper until $T = -1.5 \Gyr$; at $T = -1.05 \Gyr$, the MS then
received an impulse from an LMC encounter, which caused the spatial
bifurcation of the MS.  The MS was then given a ``kick'' by the LMC at
the subsequent apo-Galacticon at $T = -0.55 \Gyr$.  This encounter at
$T=-0.55\Gyr$ resulted in the MS being broken into two kinematic
components, resulting in the apparent velocity bifurcation.

On the other hand, a portion of the LAF comes from the particles on
the same side of the disc, but closer to the SMC centre at the $T =
-2.2 \Gyr$ encounter.  The opposite side of the edge of the disc
mostly consisted of particles that end up at the current time in the
ICR.  At $T=-0.9\Gyr$, the LMC passed through the LAF, splitting it
into two bifurcated radial components.

Our models seem to create naturally both spatial and velocity
bifurcations, via tidal interaction between the LMC and SMC.  If this
is the case, any observed bifurcation features would be strong
evidence supporting the tidal formation scenario for the MS.
Observationally, the left panel of \Fig{skymom0} shows that there is
spatial bifurcation in the observed \HI{} distribution.  However, from
this data alone, it is difficult to ascertain from the $\vlsr$
distribution of \Fig{vlsrtheta} whether there is an observed velocity
bifurcation.  Nevertheless, both \Fig{vlsrtheta} of this work, and
\citet{bks+05}, with data of higher velocity resolution, show evidence
for a bifurcation in \vlsr{} along the MS, with the two components
observed in this work being separated by approximately $100\kms$ at
$\lMC \sim 100\degr$, and two components being visible in fig.~3 of
\citet{bks+05} between the interface region and Galactic Plane.  There
is also a hint of bifurcation (with separation of $\sim 100\kms$) in
the LAF at $\lMC \sim 45\degr$.  Unfortunately, the current quality of
the observational data is not enough to lead to a firm conclusion.
Higher quality data cubes with improved velocity resolution will
provide critical information concerning this apparent bifurcated
structure.

\section{Parameter Dependences}
\label{sec:paramdep}

\begin{figure}
  \begin{center}
    \includegraphics[width=8cm]{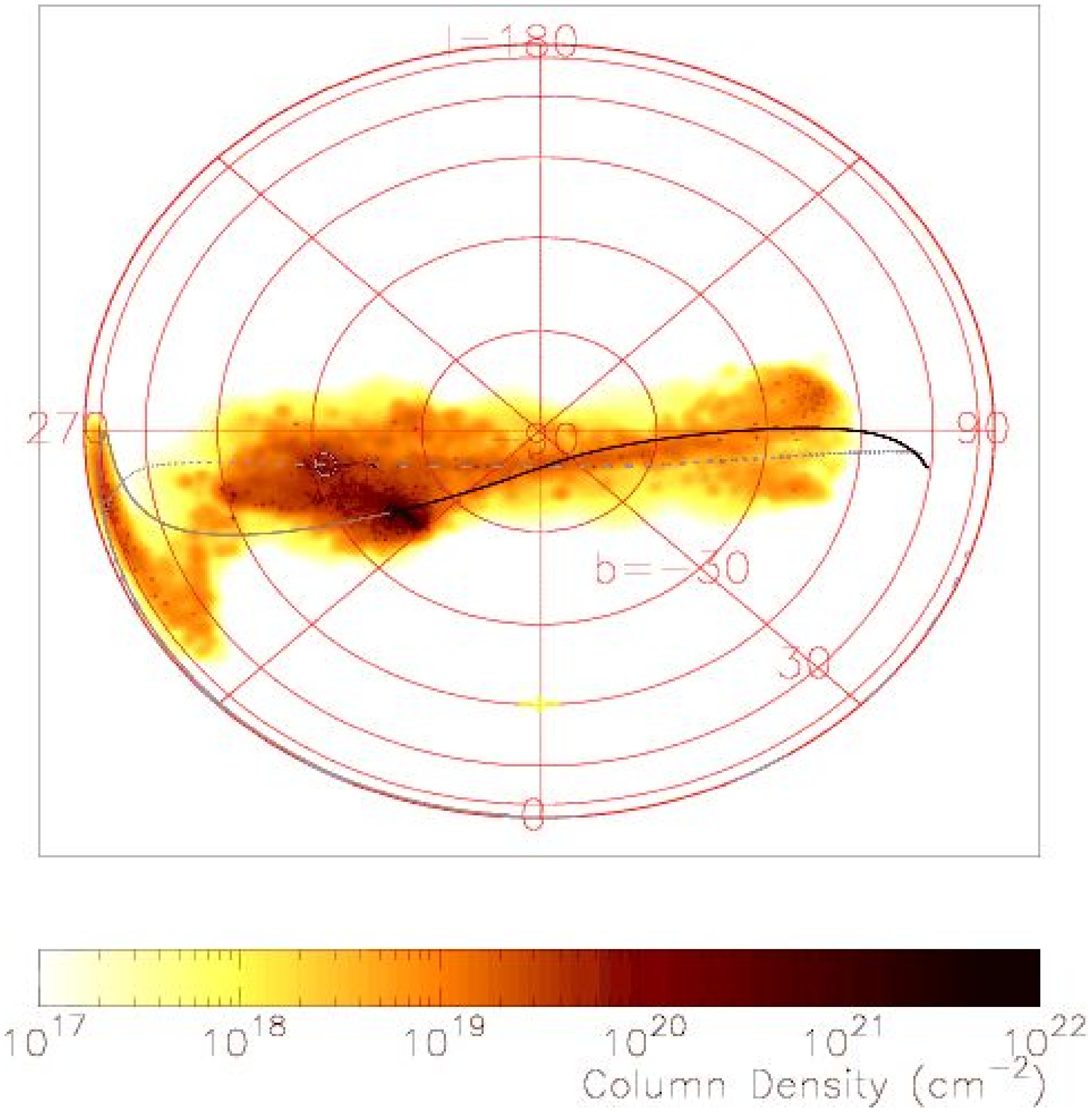}
  \end{center}
  \caption[\HI{} column density map for low-resolution fiducial
  model]{\HI{} column density map for the lower-resolution simulation
    with the same parameter set as the fiducial model.}
  \label{fig:mom0.fiducial} 
\end{figure}

\begin{figure}
  \begin{center}
    \includegraphics[width=8cm]{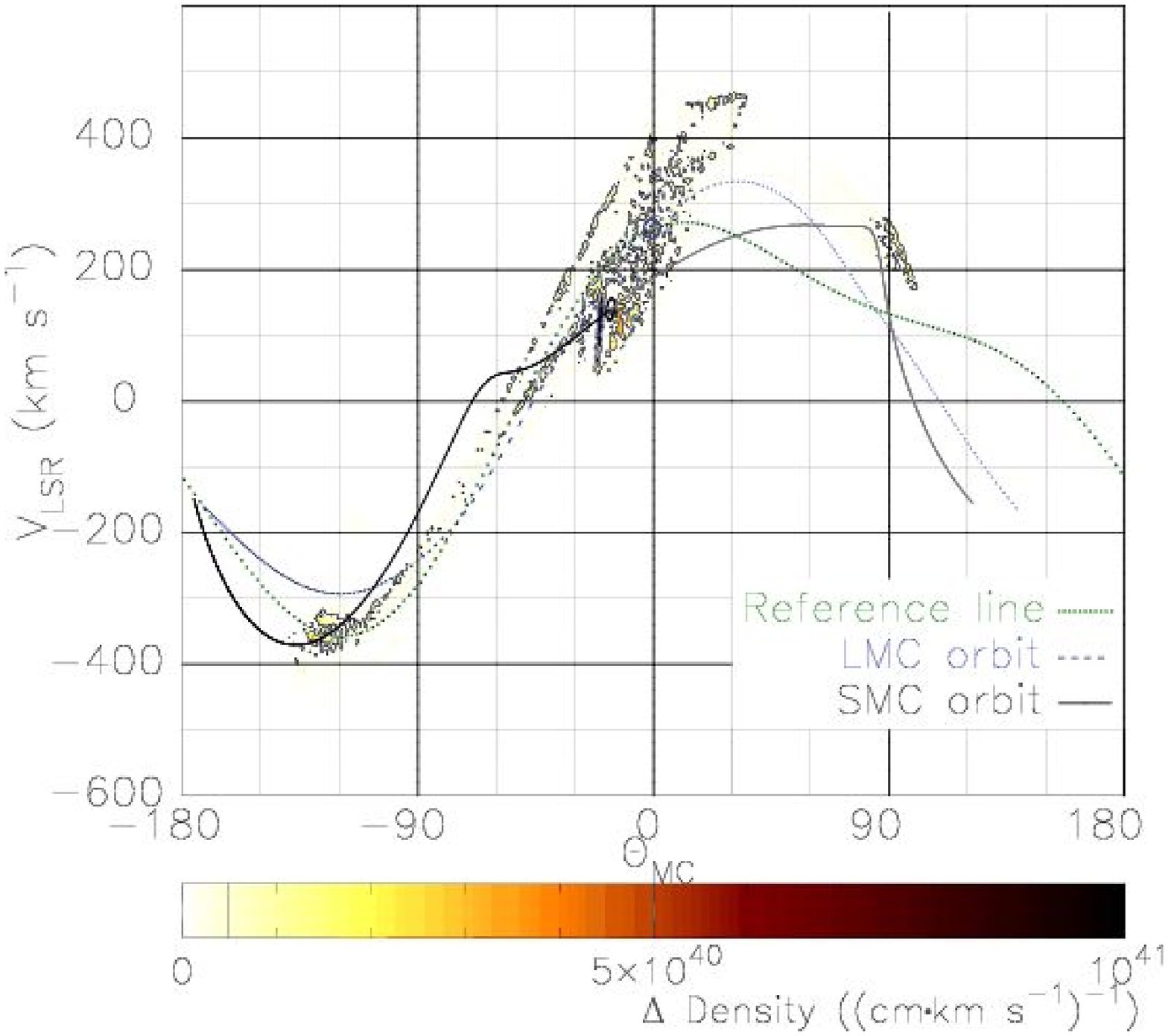}
  \end{center}
  \caption[\HI{} flux excess of high-resolution over low-resolution
  model]{\HI{} flux excess in the $\vlsr$ vs.\ $\lMC$ plane, for the
    high-resolution model over the low-resolution model, demonstrating
    the finer features visible.  Contours are plotted at $5\e{39}$ and
    $5\e{40} ({\rm cm} \kms)^{-1}$.}
  \label{fig:vlsrth_h-l} 
\end{figure}

As explained in \Sec{simulation}, our simulations involve several
parameters which are not well-constrained by current observations.  In
this section, we briefly demonstrate how the final configurations of
the MS and LAF are sensitive to these free parameters.  We varied
parameters over a wide range of parameter space in our survey, to
obtain our fiducial model.  The parameter survey was performed at a
lower resolution (25,000 disc and 25,000 halo particles) than the
final fiducial model.  We have confirmed that in the fiducial model
the results of the lower-resolution simulation are consistent with the
higher-resolution simulation, although, as presented above, the choice
of a higher resolution enables us to discuss more detailed features.
For example, in \Fig{mom0.fiducial}, we show the \HI{} column density
map for the lower-resolution model with identical parameters as our
fiducial model.  The distribution of the \HI{} column density is
roughly consistent with that shown in \Fig{skymom0}, although the
higher-resolution models affords an improved examination of the
finer-scale structures intrinsic to the simulated streams.  The
bifurcation alluded to in the previous section is an example of such a
feature, and is perhaps best appreciated through an inspection of
\Fig{vlsrth_h-l}, where the flux ``excess'' of the high-resolution
model with respect to the lower-resolution model is presented in the
$\vlsr$ vs.\ $\lMC$ plane, and the bifurcation becomes becomes
readily apparent.

Most of the parameter space surveyed is summarised in
\Tab{paramranges}.  The scale height of the SMC disc, the ratio of SMC
disc mass (\HI{} mass in \Tab{paramranges}) to SMC halo (DM) mass, the
velocity dispersion of the SMC disc, and the initial and final total
mass of the SMC (varying the mass of the SMC for the purposes of the
orbit calculation) were found not to be important to the evolution of
the system.  Both the velocity dispersion, and the ratio of \HI{} to
DM mass of the disc, simply scale the velocity dispersion and quantity
of \HI{} found in the final Streams and Clouds in a linear fashion.

In the parameter survey, the orbits of the MW, LMC and SMC were
derived assuming the mass of the SMC is constant.  However, we found
that the SMC bound mass decreases from $3\e9$ to $1.5\e9 \msun$
approximately linearly with time between $T=-2.25 \Gyr$ (near the
first interaction between the LMC and SMC), and $T=0$.  Such mass-loss
may affect the orbit and the final features of the MS and LAF (e.g.\
\citealp{z04,kgk+05}).  Thus, we explore the effects of decreasing the
mass of the SMC linearly with time from $3.0\e9$ to $1.5\e9 \msun$.
We found that there is little change to the orbits, since the LMC,
rather than the less massive SMC, primarily determines the orbit of
both bodies in \eqnsII{orbaccl}{orbaccs}.  As a result, we confirmed
that the final features of the MS and the LAF are also not influenced
heavily by the time evolution of the SMC mass.  Thus, in all other
simulations in the parameter survey, we use the orbits predicted with
no evolution of the SMC mass.  We now highlight the influence of the
most important input parameters.

\subsection{The initial scale length of the SMC disc}
\label{sec:paramdep/scalelength}

\begin{figure}
  \begin{center}
    \includegraphics[width=8cm]{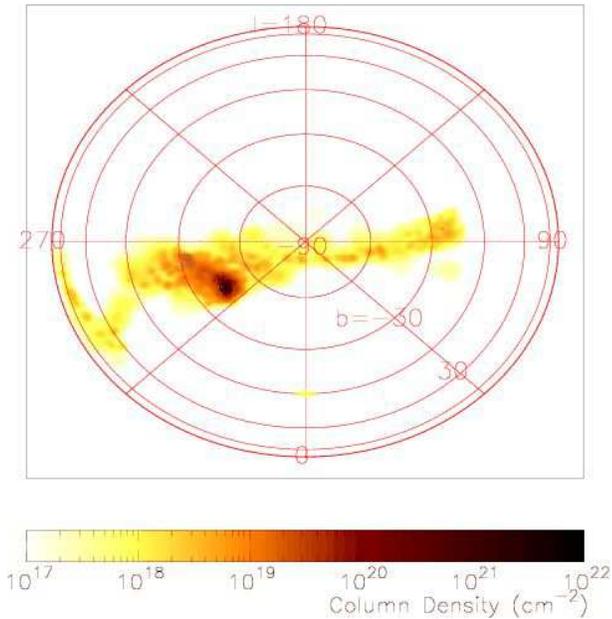}
  \end{center}
  \caption[\HI{} column density map for small scale length
  model]{\HI{} column density map for models with a smaller scale
    length ($1.4 \kpc$) for the initial SMC disc.}
  \label{fig:mom0.scalelength} 
\end{figure}

We found that the final \HI{} distributions are sensitive to the scale
length of the initial SMC disc.  \Fig{mom0.scalelength} shows the
column density map of a model with a smaller scale length ($1.4 \kpc$)
for the initial SMC disc.  This reduced scale length results in a
lower total \HI{} flux for the MS, as the initially more concentrated
SMC mass distribution results in less material being stripped from the
disc.  Since the fiducial model does not have a total \HI{} flux high
enough to match perfectly the observations
(\Sec{fiducial/HIcolumndens}), we conclude that reducing the SMC scale
length is not appropriate.  It is also worth noting that the smaller
scale length model leads to a less significant bifurcation in the
simulated MS.

To quantify the difference between the models, we have measured the
\HI{} masses and total fluxes within the regions delineated as the MS
in \Fig{skymom0}.  As a result, the low-resolution simulation of the
fiducial model has an \HI{} mass in the MS of $2.4\e8 \msun$ and total
\HI{} flux of $2.5\e5 \Jykms$.  The low-resolution simulation has a
very similar mass to that of the high-resolution case (see
\Sec{fiducial/HIcolumndens}).  On the other hand, the small scale
length model leads to a MS \HI{} mass and total flux of $1.8\e8 \msun$
and $1.9\e5 \Jykms$ respectively, which is significantly smaller than
that of the fiducial model.  Since the fiducial model has a mass which
is somewhat lower than that inferred from the HIPASS dataset, we again
conclude that the models derived with the reduced initial scale length
for the SMC perform worse than the fiducial selection.

The \HI{} survey of late-type dwarf galaxies by \citet{svv+02}
suggested that the range of scale lengths of the gas disc for galaxies
which have a similar \HI{} mass to that of the SMC is 1.5 -- $4.5
\kpc$.  The scale length of our fiducial model ($3.5 \kpc$) appears
reasonable, and thus the progenitor of the SMC is considered to have
had a large \HI{} disc before the SMC fell towards the MW.

We note that choosing a smaller scale radius is similar to setting a
small truncation radius, in that the SMC material is concentrated more
strongly towards the centre of the SMC, where it is more difficult to
tidally strip.  We obtain similar results to the above when we choose
smaller SMC truncation radii, such that the stream is substantially
retarded when the truncation radius is reduced from $7 \kpc$ to $4
\kpc$.  In this situation, the MS \HI{} mass and total flux are
reduced to $7.2\e7 \msun$ and $7.0\e4 \Jykms$.  The LAF mass is scaled
in a similar manner in both cases.

\subsection{The inclination angle of the SMC}

\begin{figure*}
  \begin{center}
    \includegraphics[width=8cm]{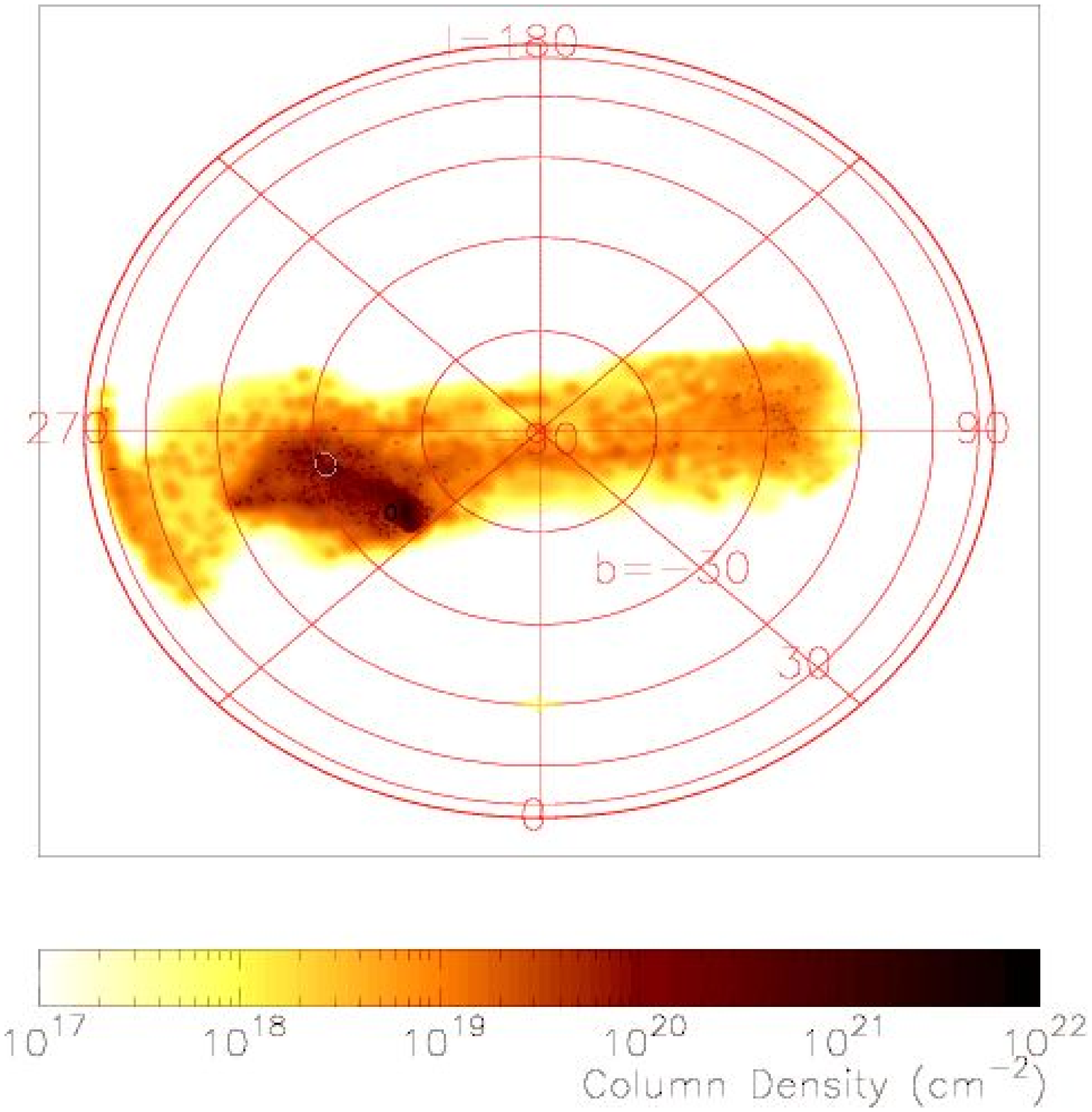}
    \includegraphics[width=8cm]{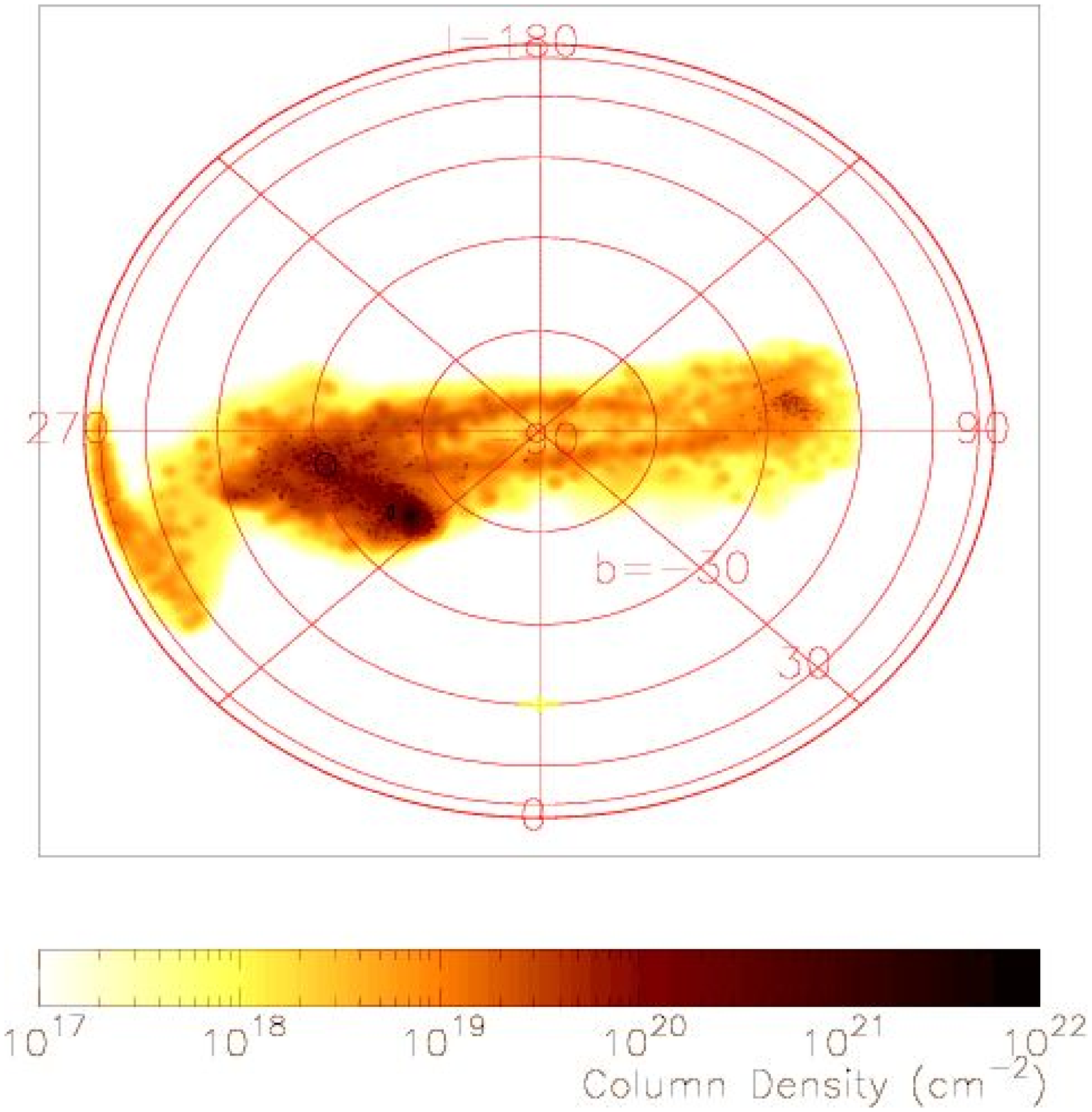}
  \end{center}
  \caption[\HI{} column density map for differing angles of SMC
  disc]{\HI{} column density map for the model $(\theta, \phi) =
    (30\degr,210\degr)$ (left) and $(45\degr,230\degr)$ (right).  The
    current positions of the SMC and LMC are represented by a black
    ellipse (the ellipse is the projection of the SMC disc at the
    given angle) and blue circle respectively.}
  \label{fig:mom0.angle} 
\end{figure*}

As mentioned in \Sec{simulation/evolve}, the current inclination angle
of the SMC is unknown, and the initial disc angle $\theta$ and $\phi$
are free parameters in our simulations.  \Fig{mom0.angle} shows the
\HI{} column density distribution for the models with $(\theta, \phi)
= (30\degr,210\degr)$ and $(45\degr,230\degr)$, to demonstrate how these
angles affect the final distribution of the MS and LAF.  We remind the
reader that we use $(\theta, \phi)=(45\degr,210\degr)$ in the fiducial
model, and therefore small differences of only $20\degr$ in the initial
inclination angle have quite a marked effect on the details of the
final distribution.  If accurate observations of the current
inclination angle of the SMC were to be made, it would provide a
strong constraint on any putative model of the formation of the MS.

\subsection{The mass of the LMC}

\begin{figure}
  \begin{center}
    \includegraphics[width=8cm]{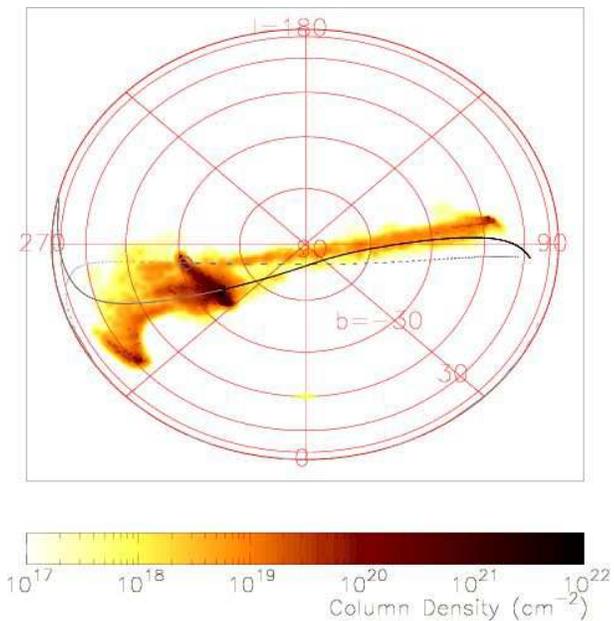}
  \end{center}
  \caption[\HI{} column density map for small LMC mass]{\HI{} column
    density map for the model with an lower LMC mass of $1.5\e{10}
    \msun$.  The past and future $1 \Gyr$ histories of the orbit
    resulting from this differing LMC mass are presented by black
    solid and blue dashed lines for the SMC and LMC respectively.}
  \label{fig:mom0.lmcmass} 
\end{figure}

Another unknown parameter is that of the mass of the LMC itself.  At
the time of the \citetalias{gn96} study, the mass of the LMC was
believed to be $\sim 2\e{10} \msun$ (e.g.\ \citealp{sso+92}).
However, recently, some authors claim the lower mass of the LMC (e.g.\ 
\citealp{vah+02} who suggested that the mass of the LMC within $R<8.9
\kpc$ is about $9\e9 \msun$).  Motivated by such claims, we also ran
models with varying orbital parameters, and in particular, a lower LMC
mass.  \Fig{mom0.lmcmass} shows the result of a model with an LMC mass
of $1.5\e{10} \msun$.  Even such small differences in the LMC mass
cause a large change in the orbits of the LMC and SMC.  Since the MS
follows the past orbit of the MCs, the angle of the MS is radically
different between the model with small LMC mass, and both the fiducial
model and the observed MS.  This result suggests that such a low mass
LMC is unlikely, {\it if} the MS is the result of tidal interactions.

\section{Summary and Discussions}
\label{sec:summary}

We have carried out high-resolution N-body simulations of the history
of the SMC disturbed by tidal interactions with the MW and LMC.  We
have surveyed most of the possible parameter space for the SMC orbit
and the properties of the initial SMC disc, and found the best model.
The increased numerical resolution of $\sim 7 \Jykms$ per particle for
the \HI{} flux, $\sim 30$ times higher than the previous studies
(\citetalias{gn96}; \citetalias{g99}; \citetalias{yn03}), made it
possible for us to examine the detailed features of both leading and
trailing tidal streams.  Taking advantage of this higher resolution,
we for the first time made a direct and quantitative comparison of the
simulation results with newly available high quality observational
data from HIPASS.  We convolved the HIPASS dataset with the identical
software tools used to analyse the simulated datasets, and compared
the results in identical manners.  Such comparisons confirm the
conclusions of previous studies -- that the existence of the LAF and
MS can be explained by the leading and trailing streams of the SMC
created by tidal interactions with the MW and LMC.  However, our
quantitative comparison have revealed extant problems with the models
(some minor; some more significant).  We found that even in our best
model (1) the shape of the MS is too extended in both width and length
(\Sec{fiducial/HIcolumndens}); (2) the total \HI{} flux of the MS is
too low, and thus the mass of the MS is too low
(\Sec{fiducial/HIcolumndens}); (3) the angle the LAF emanates from the
MCs is not entirely consistent with observations
(\Sec{fiducial/HIcolumndens}); (4) and nor is its total \HI{} flux
consistent with observations (\Sec{fiducial/HIcolumndens}); (5) the
velocity dispersion of the MS is too high (\Sec{fiducial/HImoment});
and (6) the line of sight velocity of the LAF is too high
(\Sec{fiducial/HImoment}).

These problems suggest that additional physics may be required to
explain the observed properties of the MS and LAF, quantitatively.
Obvious physics which this study excludes are gas physics, such as
hydrodynamics and dissipation by radiative cooling.  Simulations in
\citet{mkg02} indicate that dissipation causes the MS to become
narrower, which might lead to a reduced velocity dispersion.  Ram
pressure \citep{md94,mmm+05} is another physical process which our
numerical model does not currently take into account.  Ram pressure
(or drag -- see \citealp{g99}) is expected to shorten the leading arm
and increase the gas density in the MS, which should help to solve the
deficiencies of the model at recreating the LAF and MS with their
correct shapes and densities.  If it were to also bring the MS
significantly closer than the $57 \kpc$ fixed for the observational
data within this paper, then the mass of the observed stream and total
flux of the modelled stream would be increased, bringing them back
into agreement.  However, drag might lead to more problems for the
length of the MS.  Another possibly important mechanism is supernovae
feedback which might aid in ejecting gas from the SMC and/or LMC, and
help to increase the gas density in the MS and LAF.  Finally, we note
that the LAF passed very close to the centre of the MW $0.2\Gyr$ ago,
and since we modelled the MW potential as spherically symmetric with a
constant rotational velocity of $220\kms$, any deviations from this
(such as the unknown contribution from the MW disc; e.g.\ 
\citealp{ft91}) would affect our modelled LAF in particular.  We are
in the midst of introducing these physical processes into our
simulations, and will report upon their respective effects in
Paper~II.

Another benefit of the higher-resolution simulations performed here is
the identification of the bifurcation of the MS and LAF both spatially
and kinematically.  Our simulations predict that if the MS is created
by a tidal interaction with the LMC, the bifurcation would appear both
in the \HI{} column density map and the line-of-sight velocity versus
Magellanic longitude plane.  Current observations are consistent with
the existence of a spatial bifurcation in the \HI{} column density
map (left panel of \Fig{skymom0}).  In the velocity versus Magellanic
longitude plane, it is difficult to make a firm conclusion from the
current HIPASS data, although it is interesting that the SMC itself is
found to consist of at least two velocity components \citep{mf84},
perhaps caused by the same tidal disruption processes that form the
bifurcation in our models.  Observations with high velocity resolution
and sensitivity may be able to test our prediction of bifurcation
within the MS.  If confirmed, it would provide strong evidence that
the MS was created by tidal interactions.

\section*{Acknowledgements}

We would like to thank Masafumi Noguchi who kindly provided his code
for the orbit calculation, Mary Putman for the carefully flagged MS
portion of the HIPASS HVC cubes, and the anonymous referee for many
useful suggestions to improve the manuscript.  TC would like to thank
Jeremy Bailin, Chris Power, Chris Thom, Virginia Kilborn, Lister
Staveley-Smith and Geraint Lewis for helpful discussions, Stuart Gill
for his \pview{} visualisation tool source code during development,
and Christian Br\"uns for an advance copy of his paper before
publication.  The support of the Australian Research Council and the
Australian and Victorian Partnership for Advanced Computing, the
latter through its Expertise Grant Program, is gratefully
acknowledged.  DK acknowledges the financial support of the JSPS,
through Postdoctoral Fellowship for research abroad.  In this work we
make use of data from HIPASS, which is operated by the Australia
Telescope National Facility, CSIRO.

\bibliography{ckgms_astroph}

\label{lastpage}
\end{document}